\documentclass[12pt]{article}
\usepackage{graphicx}
\usepackage{amsmath}
\usepackage{amsfonts}
\usepackage{amssymb}
\newtheorem{theorem}{Theorem}

\newtheorem{corollary}[theorem]{Corollary}

\newtheorem{definition}[theorem]{Definition}

\newtheorem{lemma}[theorem]{Lemma}

\newtheorem{proposition}[theorem]{Proposition}
\newtheorem{remark}[theorem]{Remark}

\newenvironment{proof}[1][Proof]{\textbf{#1.} }{\ \rule{0.5em}{0.5em}}
\newcommand{\rem}[1]{}

\begin{document}
\date{October 25, 2009}
\title{Stochastic attractors for shell phenomenological
models of turbulence}

\author{Hakima Bessaih\thanks{University of Wyoming, Department of Mathematics, Dept.
3036, 1000 East University Avenue, Laramie WY 82071, USA,
bessaih@uwyo.edu}, Franco Flandoli\thanks{ Dipartimento di
Matematica applicata ``U. Dini", Universit\`a di Pisa, Via
Buonarrotti 1, 56127 Pisa, Italy, flandoli@dma.unipi.it} , Edriss S.
Titi\thanks{Department of Mathematics and Department of Mechanical
and Aerospace Engineering, University of California, Irvine CA
92697,USA, etiti@math.uci.edu, {\bf also} Department of Computer
Science and Applied Mathematics, The Weizmann Institute of Science,
Rehovot 76100, Israel}}


\maketitle


\begin{abstract} Recently, it has been proposed that the Navier-Stokes
equations and a relevant linear advection model  have the same
long-time statistical properties, in particular, they have the same
scaling exponents of their structure functions. This assertion has
been investigate rigorously in the context of certain nonlinear
deterministic phenomenological shell model, the Sabra shell model,
of turbulence and its corresponding linear advection counterpart
model.  This relationship has been established through a
``homotopy-like" coefficient $\lambda$ which bridges continuously
between the two systems. That is, for $\lambda=1$ one obtains the
full nonlinear model, and the corresponding linear advection  model
is achieved for $\lambda=0$. In this paper, we investigate the
validity of this assertion for certain stochastic phenomenological
shell models of turbulence driven by an additive noise. We prove the
continuous dependence of the solutions with respect to the parameter
$\lambda$. Moreover, we show  the existence of a finite-dimensional
random attractor for each value of $\lambda$ and establish the upper
semicontinuity property of this random attractors, with respect to
the parameter $\lambda$. This property is proved by a pathwise
argument. Our study aims toward the development of basic results and
techniques that may contribute to the understanding of the relation
between the long-time statistical properties of  the nonlinear and
linear models.

\end{abstract}
\section{Introduction}

\subsection{Motivation}

The GOY shell model \cite{Gledzer} and \cite {Yamada-Ohkitani}, and
Sabra shell model \cite{Lvov} are some of the most interesting and
most popular examples of simplified phenomenological models of
turbulence. This is because, although departing from reality, they
capture some essential statistical properties and features of
turbulent flows, like the energy and the enstrophy cascade, and the
power law decay of the structure functions in some range of wave
numbers - the inertial range. We refer the reader to, e.g.,
\cite{procaccia2001}, \cite{biferale}, \cite{procaccia2002},
 \cite{constantin-titi}, \cite{Fri},
 and references therein for several descriptions
and results. Often, in numerical or theoretical investigations, such
models are driven by white noise forces. Both the stochastic GOY and
Sabra shell models  have the form
\begin{equation}
du_{n}+\left(  \nu k_{n}^{2}u_{n}+b_{n}\left(  u,u\right)  \right)
dt=\sigma_{n}d\beta_{n},\quad n=1,2,...\label{GOY}%
\end{equation}
where $u_{n}\left(  t\right)  $ are complex valued, $\nu>0$ is a
parameter that represents the viscosity, $k_{n}=k_{0}2^{n}$ for some
$k_{0}>0$ are representing wave numbers, $u(t)$ denotes the sequence
$\left( u_{n}\left( t\right)  \right)  _{n\geq1}$, $b_{n}\left(
\cdot,\cdot\right)  $ is a complex valued bilinear function of
complex sequences $u=\left(  u_{j}\right)  _{j\geq1}$, that depends
depending only on the variables $u_{n-2},u_{n-1},u_{n+1},u_{n+2}$
(where we impose the boundary conditions $u_{-1}\left(  t\right)
=u_{0}\left( t\right) =0$). $\sigma_{n}$ is a sequence of complex
numbers, that are usually chosen equal to zero for all $n$ greater
than some $n_{0}$ (which describes the range of wavenumbers and
consequently the length scales of external forces), $\left(
\beta_{n}\right) _{n\geq1}$ is a sequence of independent complex
valued Brownian motions. A rigorous theoretical analysis of the
stochastic equation \eqref{GOY} and some of its statistical
properties have been investigated in \cite{flandoli-goy}, while
other rigorous results in the case of deterministic force have been
developed in \cite{BLPTiti}, \cite{constantin-titi},
\cite{constantin-titi2}, \cite{constantin-titi3}. The exact form of
$b_{n}\left( \cdot,\cdot\right) $ varies from one model to  another.
However, in all the various models in the sequel we assume that
$b_{n}\left( \cdot,\cdot\right) $ is chosen in such a way that
\begin{equation}\label{energy}
\sum_{n=1}^{\infty}b_{n}\left(  u,v\right)  \overline{v}_{n}=0,
\end{equation}
for all square summable sequences $u=\left(  u_{m}\right) _{m\geq1}$
and $v=\left(  v_{m}\right)  _{m\geq1}\,$. Equation~(\ref{energy})
implies a formal law of the conservation of energy in the inviscid
($\nu=0$) and unforced form of~(\ref{GOY}).

In analogy with the statistical theory of turbulence it is
interesting to investigate the accompanying linear advection
equation to equation~(\ref{GOY}), that is the linear auxiliary
linear equation in the unknown $w\left( t\right) =\left( w_{n}\left(
t\right)  \right) _{n\geq1}$
\begin{equation}
dw_{n}+\left(  \nu k_{n}^{2}w_{n}+b_{n}\left(  u,w\right)  \right)
dt=\sigma_{n}d\beta_{n},\quad n=1,2,...\label{linear_GOY}%
\end{equation}
where $u$ is the solution of~(\ref{GOY}), and $w_{n}\left(  t\right)
$ are  complex valued functions. There is an extensive literature
investigating the statistical properties of linear advection
(passive-scalar) equations in turbulent flows, which we do not
pretend to cover in this contribution. We observe, however, that
equation~(\ref{linear_GOY}) is not the linearized version of
equation of (\ref{GOY}) about the solution $u$. This is because the
term $b_{n}\left( w,u\right)  $ is missing from~(\ref{linear_GOY}),
and an additive force still appears in the right-hand side
of~(\ref{linear_GOY}). Equation (\ref{linear_GOY}) should be
considered as an auxiliary equation which, to some extent, may have
 similar statistical properties to those of equation (\ref{GOY}), but
 is amenable to linear analysis (for instance the use of propagators). There is
some numerical and heuristic evidence that some statistical
properties of the solutions to equation (\ref{linear_GOY}), like the
scaling exponents of the structure functions, are the same as those
of the solutions to equation (\ref{GOY}), see \cite{ABBPT} and
\cite{BLPTiti}. It is then of interest to understand the properties
of the joint system
\begin{align}
du_{n}+\left(  \nu k_{n}^{2}u_{n}+b_{n}\left(  u,u\right)  \right)  dt &
=\sigma_{n}d\beta_{n}\label{system0}\\
dw_{n}+\left(  \nu k_{n}^{2}w_{n}+b_{n}\left(  u,w\right)  \right)
dt & =\sigma_{n}d\beta_{n}\, ,\nonumber
\end{align}
for $n=1,2,...$  In addition, the following idea has been introduced
first in \cite{ABBPT} and proved rigorously later in \cite{BLPTiti}:
one can  symmetrize  system~(\ref{system0}) by means of two
additional terms as follows
\begin{align}
du_{n}+\left(  \nu k_{n}^{2}u_{n}+b_{n}\left(  u,u\right)  +\lambda
b_{n}\left(  w,u\right)  \right)  dt &  =\sigma_{n}d\beta_{n}%
\label{final_system}\\
dw_{n}+\left(  \nu k_{n}^{2}w_{n}+b_{n}\left(  u,w\right)  +\lambda
b_{n}\left(  w,w\right)  \right)  dt & =\sigma_{n}d\beta_{n}\,
,\nonumber
\end{align}
where $\lambda\in \mathbb{R}$ is a parameter, and to analyze the
dependence on $\lambda$ of the properties of~(\ref{final_system}).
For $\lambda=0$ we recover (\ref{system0}). Observe that for
$\lambda\neq 0$, setting $v=\lambda w$ and multiplying the second
equation by $\lambda$, we have a perfectly symmetric system for the
pair $\left( u,v\right) $, except for the force and initial
conditions. Thus, to some extent, we would expect that $u$ and
$\lambda w$ have similar statistical properties for $\lambda\neq 0$.
If we consider, for instance, the structure function $S_{p}\left(
k_{n}\right) =\left\langle \left| u_{n}\right| ^{p}\right\rangle $
(we do not specify at this heuristic level the meaning of the
averaging procedure $\left\langle .\right\rangle $), one might
expect, as in the case of turbulent flows, that
\[
S_{p}\left(  k_{n}\right)  \sim k_{n}^{-\zeta_{p}}%
\]
for $n$ lies in the so-called inertial range. The numbers
$\zeta_{p}$ are called scaling exponents, which are universal in
turbulent flows as the Reynolds number tends to infinity, i.e., as
the viscosity tends to zero. Therefore, one possible definition of
$\zeta_{p}$ is
\[
\lim_{\left(  \nu,k_{n}\right)  \rightarrow\left(  0,\infty\right)  }%
\frac{\log S_{p}\left(  k_{n}\right)  }{\log k_{n}}=-\zeta_{p},%
\]
where the limit is taken along a region of the form
$\nu^{\alpha}\leq k_{n}^{-1}\leq\nu^{\beta}$ for some $\alpha>0$
(usually $\alpha=\frac{4}{3}$). This is in order to ensure that the
wavenumber considered are lying in the heart of the inertial range,
as the viscosity tends to zero. Such statistical property,
\textit{if} it holds for $w$, it holds as well for $\lambda
w$ with the same value $\zeta_{p}$ (and vice versa): indeed, if $S_{p}%
^{\left(  w\right)  }\left(  k_{n}\right)  $ and $S_{p}^{\left(
\lambda w\right)  }\left(  k_{n}\right)  $ are the structure
functions of $w$ and $\lambda w$, respectively, we have
$S_{p}^{\left(  \lambda w\right)  }\left( k_{n}\right)
=\lambda^{p}S_{p}^{\left(  w\right)  }\left(  k_{n}\right)  $ and
$\lim_{\left(  \nu,k_{n}\right)  \rightarrow\left( 0,\infty\right)
}\frac{\log\lambda^{p}}{\log k_{n}}=0$, which imply the claim. Thus,
\textit{if} the scaling exponents $\zeta_{p}$ exist for both $u$ and
$\lambda w$ (this assumption seems to be reasonable based on
numerical finding in \cite{ABBPT}, \cite{BLPTiti} and the references
therein) and are equal (which is reasonable to assume thanks to the
symmetry $u\leftrightarrow\lambda w$ described above), then they are
equal for $u$ and $w$. In summary, it is, therefore, reasonable to
expect that some statistical properties like the existence and the
value of scaling exponents, are the same for $u$ and $w$, whenever
$\lambda\neq 0$.

Finally, it will be of great interest to show that such statistical
properties depend continuously on $\lambda$, as
$\lambda\rightarrow0$: if this is true, then the solutions of
(\ref{GOY}) and (\ref{linear_GOY}) have the same statistical
properties of the kind just described above. In particular, if this
program is true, one is sure that results for the simpler linear
model (\ref{linear_GOY}) can be translated to (\ref{GOY}), which
will be a remarkable breakthrough.

\subsection{Content of the paper}

The program above, outlined in \cite{ABBPT} and \cite{BLPTiti}, is
composed of several steps, some of them are not easy to be justified
rigorously. The first rigorous result has been obtained in
\cite{BLPTiti} states that: in the case of deterministic forces,
solutions of (\ref{final_system}) depend continuously on $\lambda$
in $C\left( \left[ 0,T\right] ;H\times H\right)  $, for every given
$T>0$. Here $H$ is the space of square summable sequences $\left(
v_{n}\right) _{n\geq1}$ in $\mathbb{C}$. This implies that the
structure functions, defined as time average on any fixed finite
time interval $\left[ 0,T\right] $:
\[
S_{p}\left(  k_{n}\right)  =\frac{1}{T}\int_{0}^{T}\left(  \left|
u_{n}\left(  t\right)  \right|  ^{p}\right)  dt
\]
depend continuously on $\lambda$. One of the limitations of this
result of  \cite{BLPTiti} that it  considers deterministic forces.
Here, we remove this restriction and prove the same result in the
case described above of white noise forces.

Several other issues have to be solved in order to be able to claim
that the program described above is complete. One of the other major
issues  in \cite{BLPTiti} is that the statistics is being considered
on finite intervals of time $[0,T]$ instead of being considered on
the attractor, i.e. as $T\longrightarrow\infty$. The existence of
the limit as $T\rightarrow\infty$, in the time average (definition
of $S_{p}\left(  k_{n}\right)  $) of the deterministically forced
system is, therefore, one of these issues of \cite{BLPTiti}. We do
not directly address this difficult problem here, in the
stochastically forced case,  but we content ourselves with a
structural result about the infinite time horizon properties of
(\ref{final_system}): we prove existence of a finite-dimensional
random attractor. This is a pathwise property, in the vein of the
property of continuous dependence on $\lambda$ in $C\left(  \left[
0,T\right]  ;H\right) $ stated above for the deterministic case. We
hope that this result, or the techniques involved in establishing
it, may contribute to the understanding of the problem of the
long-term behavior, i.e. $T\rightarrow\infty$. Notice that we
construct the random attractor for system (\ref{final_system}) for
every $\lambda\in \mathbb{R}$, hence if we take in particular
$\lambda=0$ the first component of the system is decoupled and thus
the projection of the attractor on the first component is the random
attractor of equation (\ref{GOY}). Thus  we prove in this paper the
existence of a finite-dimensional random attractor for the
stochastic GOY and Sabra shell models, as a particular case of a
more general result. However, the general result for system
(\ref{final_system}) may help to prove further results on the
relations between the statistics of the  nonlinear and the linear
cases.

Due to the It\^{o} nature of the previous equations, it is clear
that other kind of analysis could be performed, in distribution
and average sense instead of pathwise. This will be done
elsewhere. We restrict ourselves here to purely pathwise
properties.

The paper is organized as follows. In section 2, we present the
functional framework and prove pathwise well-posedness of system
(\ref{final_system}), and the continuous dependence of the solutions
on $\lambda$. In section 3, we give some preliminary results about
random attractors and some of their properties. In section 4, we
prove the existence of a random attractor for every coefficient
$\lambda \in \mathbb{R}$, its upper semicontinuity with respect to
$\lambda$; and finally that the random attractor has a finite
Hausdorff dimension.

\section{Well-posedness and continuous dependence on $\lambda$}

\subsection{Functional setting}

Let us introduce the following spaces of complex valued sequences;
we consider them as vector spaces on the field of real numbers.
The space $H$ is the space of $l^{2}$ sequences over the field of
complex numbers $\mathbb{C}$:
\[
H=\left\{  u=\left(  u_{n}\right)  _{n\geq1}:u_{n}\in\mathbb{C}\text{ for all
}n\geq1\text{ and }\sum_{n=1}^{\infty}\left|  u_{n}\right|  ^{2}%
<\infty\right\}  .
\]
It is a Hilbert space with the inner product
\[
\left\langle u,v\right\rangle _{H}:=\operatorname{Re}\sum_{n=1}^{\infty}%
u_{n}\overline{v}_{n}%
\]
and the norm given by $\left|  u\right|  _{H}^{2}=\sum_{n=1}^{\infty}\left|
u_{n}\right|  ^{2}$. Let us recall that we have defined $k_{n}=2^{n}k_{0}$,
$n\geq1$, with $k_{0}>0$ given. We introduce now the Hilbert spaces
$D(A)\subset V\subset H$ defined as
\[
V=\left\{  u\in H:\sum_{n=1}^{\infty}k_{n}^{2}\left|  u_{n}\right|
^{2}<\infty\right\}
\]
with norm $\left\|  u\right\|
_{V}^{2}=\sum_{n=1}^{\infty}k_{n}^{2}\left| u_{n}\right|  ^{2}$.
Moreover, for all $\alpha \ge 0$, we define
\[
D(A^\alpha)=\left\{  u\in H:\sum_{n=1}^{\infty}k_{n}^{4\alpha}\left|
u_{n}\right| ^{2}<\infty\right\}  .
\]
On the latter space we define the linear operator
$A^\alpha:D(A^\alpha)\subset H\rightarrow H$ as
\[
\left(  A^\alpha u\right)  _{n}=k_{n}^{2\alpha}u_{n},\quad {\mbox
{for
 all}} \quad u\in D(A^\alpha).
\]
The operator $A^\alpha$ is self-adjoint and strictly positive
definite:
\[
\left\langle A^\alpha u,u\right\rangle _{H}\geq k_{0}^{2\alpha}\left|  u\right|  _{H}%
^{2},\quad {\mbox{for all}} \quad  u\in D(A^\alpha).
\]
We also observe that the inclusion maps of $D(A)\subset V$ and
$V\subset H$ are compact embeddings. We finally introduce the
bilinear operator $B\left( \cdot , \cdot \right) :V\times
V\rightarrow H$. For the GOY shell model it is defined as
\begin{align*}
b_n (u,v) &:= \left(B  (u,v) \right)  _{n} \\
& :=ik_{n}\left(  \frac{1}{4}\overline{v}_{n-1}%
\overline{u}_{n+1}-\frac{1}{2}\left(  \overline{u}_{n+1}\overline{v}%
_{n+2}+\overline{v}_{n+1}\overline{u}_{n+2}\right)
+\frac{1}{8}\overline {u}_{n-1}\overline{v}_{n-2}\right)  .
\end{align*} For the Sabra shell model we define it as
\begin{align*}
b_n (u,v) := &\left(B  (u,v) \right)  _{n}:=
=\frac{i}{3}k_{n+1}\left[  \left( 1+\delta\right)
\overline{v}_{n+1}u_{n+2}+\left(  2-\delta\right)
\overline{u}_{n+1}v_{n+2}\right] \\
&  +\frac{i}{3}k_{n}\left[  \left(  1-2\delta\right)  \overline{u}%
_{n-1}v_{n+1}-\left(  1+\delta\right)  \overline{v}_{n-1}u_{n+1}\right] \\
&  +\frac{i}{3}k_{n-1}\left[  \left(  2-\delta\right)  u_{n-1}v_{n-2}+\left(
1-2\delta\right)  u_{n-2}v_{n-1}\right]
\end{align*}
(see \cite{constantin-titi}, \cite{constantin-titi2},
\cite{constantin-titi3}), where $\delta$ is a real number. In both
shell models we impose the boundary conditions $u_{-1}=u_{0}=0$.
What distinguishes the Sabra shell model from the GOY one is the
dependence of the former on the parameter $\delta$, which is in
charge for changing its  character from the so-called 2d Turbulence
regime to the 3d Turbulence regime, depending on the definiteness of
the sign of a second (in addition to the energy) quadratic conserved
quantity; see \cite{BLPTiti}, \cite{constantin-titi},
\cite{constantin-titi2}, \cite{constantin-titi3} and \cite{Lvov}.

For both the GOY and Sabra shell models, the operator $B\left(
.,.\right)  $ is a bilinear continuous operator from $V\times H$
to $H$, and also from $H\times V$ to $H$, as it will be stated in
the next lemma. We also state its basic skew-symmetry property.

\begin{lemma}
\label{lemma estimate}There is a constant $C>0$ such that
\[
\left|  B\left(  u,v\right)  \right|  _{H}\leq C\left\|  u\right\|
_{V}\left|  v\right|  _{H},\quad u\in V,v\in H
\]
and
\[
\left|  B\left(  u,v\right)  \right|  _{H}\leq C\left\|  v\right\|
_{V}\left|  u\right|  _{H},\quad v\in V,u\in H.
\]
Hence, $B\left(  .,.\right)  $ is a bilinear continuous operator
from $V\times H $ to $H$, and from $H\times V$ to $H$. Moreover,
\[
\left\langle B\left(  u,v\right)  ,v\right\rangle _{H}=0\text{ }%
\]
for all $u\in V$ and $v\in H$, or $v\in V$ and $u\in H$.
Equivalently, we have
\[
\left\langle B\left(  u,v\right)  ,w\right\rangle _{H}=-\left\langle B\left(
u,w\right)  ,v\right\rangle _{H}%
\]
for all $u\in V$ and $v,w\in H$, or $v,w\in V$ and $u\in H$.
\end{lemma}

\begin{proof}
The first inequality follows from the fact that
\[
\sum_{n=1}^{\infty}k_{n}^{2}\left|  u_{n}\right|  ^{2}\left|  v_{n}\right|
^{2}\leq\left(  \sup_{n}k_{n}^{2}\left|  u_{n}\right|  ^{2}\right)  \sum
_{n=1}^{\infty}\left|  v_{n}\right|  ^{2}\leq\left\|  u\right\|  _{V}%
^{2}\left|  v\right|  _{H}^{2}%
\]
and the second inequality follows similarly by interchanging $u$ and
$v$. Having proved these facts, the expressions in the last two
identities are all well defined. It is sufficient to prove the first
identity, since it implies the second one because, if $u\in V$ and
$v,w\in H$ or $v,w\in V$ and $ u\in H$, from the first identity and
by the bilinearity of $B$ we have
\begin{align*}
0 &  =\left\langle B\left(  u,v+w\right)  ,v+w\right\rangle _{H}\\
&  =\left\langle B\left(  u,v\right)  ,w\right\rangle
_{H}+\left\langle B\left(  u,w\right)  ,v\right\rangle _{H}\,,%
\end{align*}
where we have used the fact that $\left\langle B\left(  u,v\right)
,v\right\rangle _{H}=\left\langle B\left(  u,w\right)
,w\right\rangle _{H}=0$, by the first identity again. This implies
the second one. One can also prove the converse.

Finally, let us prove that
\[
\sum_{n=1}^{\infty}\operatorname{Re}\left[  B\left(  u,v\right)  _{n}%
\overline{v}_{n}\right]  =0.
\]
For the GOY model we have
\begin{align*}
&  \frac{-i}{k_{0}}\left\langle B(u,v),v\right\rangle _{H}\\
&  =\sum_{n=1}^{\infty}2^{n-2}\overline{v}_{n-1}\overline{v}_{n}\overline
{u}_{n+1}-\sum_{n=1}^{\infty}2^{n-1}\overline{v}_{n}\overline{u}%
_{n+1}\overline{v}_{n+2}\\
&  -\sum_{n=1}^{\infty}2^{n-1}\overline{v}_{n}\overline{v}_{n+1}\overline
{u}_{n+2}+\sum_{n=1}^{\infty}2^{n-3}\overline{v}_{n-2}\overline{u}%
_{n-1}\overline{v}_{n}\\
&  =\sum_{n=0}^{\infty}2^{n-1}\overline{v}_{n}\overline{v}_{n+1}\overline
{u}_{n+2}-\sum_{n=1}^{\infty}2^{n-1}\overline{v}_{n}\overline{u}%
_{n+1}\overline{v}_{n+2}\\
&  -\sum_{n=1}^{\infty}2^{n-1}\overline{v}_{n}\overline{v}_{n+1}\overline
{u}_{n+2}+\sum_{n=-1}^{\infty}2^{n-1}\overline{v}_{n}\overline{u}%
_{n+1}\overline{v}_{n+2}=0.
\end{align*}
The computation for the Sabra model is very similar, see, e.g.,
\cite{constantin-titi}. The proof is complete.
\end{proof}

We will also consider also the space $V^{\prime}$, the dual space of
$V$, which can be identified as
\[
V^{\prime}=\left\{  u=\left(  u_{n}\right)
_{n\geq1}:u_{n}\in\mathbb{C}\text{ for all }n\geq1, \text{ and
}\sum_{n=1}^{\infty}k_{n}^{-2}\left|  u_{n}\right|
^{2}<\infty\right\}
\]
with the norm $\left|  u\right|  _{V^{\prime}}^{2}:=\sum_{n=1}^{\infty}%
k_{n}^{-2}\left|  u_{n}\right|  ^{2}$, $u\in V^{\prime}$. It is
clear that $H\subset V^{\prime}$, and \ $V^{\prime}$ is the dual of
$V$ (with respect to $H$), with dual pairing between $V^{\prime}$
and $V$ defined as
\[
\left\langle u,v\right\rangle _{V^{\prime},V}:=\operatorname{Re}\sum
_{n=1}^{\infty}u_{n}\overline{v}_{n},\quad\forall u\in V^{\prime},v\in V.
\]
Observe that $\left\langle u,v\right\rangle _{H}=\left\langle
u,v\right\rangle _{V^{\prime},V}$, when $u\in H$, for every $v\in
V$.

It is easy to extend the operator $A$ as a bounded linear operator
from $V$ to $V^{\prime} $. One can also extend $B$ to a bilinear
operator $B\left(  .,.\right)
:H\times H\rightarrow V^{\prime}$. The definition
is possible because%

\begin{align}\label{V-prime}
\left| B(u,v)\right|^2_{V^{\prime}}=&
\sum_{n=1}^{\infty}k_{n}^{-2}\left| B(u,v)_{n}\right|  ^{2}   \leq
C_* \left(\sum_{n=1}^{\infty}|{v}_{n}|^2\right)
\left (\sup_{n \ge 1}|{u}_{n}|^2\right ) \nonumber \\
  \leq & C_*\left(  \sum_{n=1}^{\infty}|v_{n}|^{2}\right) \left(
\sum_{n=1}^{\infty}|u_{n}|^{2}\right) =C_*|u|^2_{H}|v|^2_{H}.
\end{align}
We also have
\[
\left\langle B\left(  u,v\right)  ,z\right\rangle _{V^{\prime},V}%
=-\left\langle B\left(  u,z\right)  ,v\right\rangle _{H}%
\]
for all $u,v\in H$, $z\in V$. Indeed, the identity is true for $u\in
H$,  and $v,z\in V$, because in such a case $\left\langle B\left(
u,v\right) ,z\right\rangle _{V^{\prime},V}=\left\langle B\left(
u,v\right) ,z\right\rangle _{H}$ and we may use Lemma \ref{lemma
estimate}. Then we extend the result to $v\in H$ by density of $V$
in $H$.

Define
\[
\widetilde{H}=H\times H,\quad\widetilde{V}=V\times V\text{ and }
\quad D(\widetilde{A}^\alpha)=D(A^\alpha)\times D(A^\alpha),
\]
for $\alpha \ge 0$.

If $x=(x_{1},x_{1})\in\widetilde{H}$ and
$y=(y_{1},y_{2})\in\widetilde{H}$, we define the scalar product in
$\widetilde{H}$ as
\[
<x,y>_{\widetilde{H}}=<x_{1},y_{1}>_{H}+<x_{2},y_{2}>_{H}%
\]
and the norms in $\widetilde{H}$ and $\widetilde{V}$ as
\[
|x|_{\widetilde{H}}^{2}=|x_{1}|_{H}^{2}+|x_{2}|_{H}^{2},\quad x=(x_{1}%
,x_{2})\in\widetilde{H}%
\]%
\[
\Vert x\Vert_{\widetilde{V}}^{2}=\Vert x_{1}\Vert_{V}^{2}+\Vert x_{2}\Vert
_{V}^{2}\quad x=(x_{1},x_{2})\in\widetilde{V}.
\]
Moreover, define the linear operator $\widetilde{A}:D( \widetilde
{A}) \subset\widetilde{H}\rightarrow\widetilde{H}$, or also
$\widetilde{A}:\widetilde{V}\rightarrow\widetilde{V}^{\prime}$, as
$\widetilde{A}x=\left(  Ax_{1},Ax_{2}\right)  $ and, for every
$\lambda\in \mathbb{R}$, define the bilinear continuous operator
$\widetilde{B}_{\lambda}$ from $\widetilde{V}\times\widetilde{H}$ to
$\widetilde{H}$ or from $\widetilde {H}\times\widetilde{V}$ to
$\widetilde{H}$ as
\[
\widetilde{B}_{\lambda}\left(  x,y\right)  =\left(  B\left(  x_{1}%
,y_{1}\right)  +\lambda B\left(  x_{2},y_{1}\right)  ,B\left(  x_{1}%
,y_{2}\right)  +\lambda B\left(  x_{2},y_{2}\right)  \right),%
\]
where as usual we have used the notation $x=\left(
x_{1},x_{2}\right)  $, $y=\left(  y_{1},y_{2}\right)  $. The main
properties of the operator $\widetilde{B}_{\lambda}$ are listed in
the following lemma, whose proof is an easy consequence of Lemma
\ref{lemma estimate}.

\begin{lemma}\label{lemma su B tilde}
There is a constant $C>0$ such that
\[
\left|  \widetilde{B}_{\lambda}\left(  u,v\right)  \right|  _{\widetilde{H}%
}\leq C\left\|  u\right\|  _{\widetilde{V}}\left|  v\right|  _{\widetilde{H}%
},\quad \text{for every} \quad u\in\widetilde{V},v\in\widetilde{H}\, ,%
\]
and
\[
\left|  \widetilde{B}_{\lambda}\left(  u,v\right)  \right|  _{\widetilde{H}%
}\leq C\left\|  v\right\|  _{\widetilde{V}}\left|  u\right|  _{\widetilde{H}%
},\quad \text{for every} \quad v\in\widetilde{V},u\in\widetilde{H}.
\]
Moreover,
\[
\left\langle \widetilde{B}_{\lambda}\left(  u,v\right)  ,v\right\rangle
_{\widetilde{H}}=0\, , \text{ }%
\]
for all $u\in\widetilde{V},v\in\widetilde{H}$;  or $v\in\widetilde{V}%
,u\in\widetilde{H}$;\ also,
\[
\left\langle \widetilde{B}_{\lambda}\left(  u,v\right)  ,w\right\rangle
_{\widetilde{H}}=-\left\langle \widetilde{B}_{\lambda}\left(  u,w\right)
,v\right\rangle _{\widetilde{H}}%
\]
for all $u\in\widetilde{V}$ and $v,w\in\widetilde{H}$, or $v,w\in\widetilde{V}$%
 and $u\in\widetilde{H}$.
\end{lemma}

\subsection{Well-posedness, stochastic flow and pathwise version in $\lambda$}

Let $\left(  \sigma_{n}\right)  $ be a sequence of complex numbers such that
\begin{equation}
\sum k_{n}^{2\varepsilon}\left|  \sigma_{n}\right|  ^{2}<\infty
\label{assumption on Q}%
\end{equation}
for some $\varepsilon>0$. This is a standing assumption for the sequel.

Let $\Omega$ be the space of continuous functions from $\mathbb{R}$
to $H$, null at zero, endowed with the metric of uniform convergence
on compact sets. Let $\mathcal{F}$ be the Borel $\sigma$-field
associated with $\Omega$. Denote by $\left( W\left(  t\right)
\right) _{t\in\mathbb{R}}$ the canonical process defined on $\Omega$
as $W\left( t,\omega\right)  =\omega\left( t\right) $, for every
$\omega\in\Omega$. Let $P$ be a probability measure on $\left(
\Omega,\mathcal{F}\right)  $ such that $\left(  W\left( t\right)
\right)  _{t\geq0}$ and $\left( W\left(  -t\right)  \right)
_{t\geq0}$ are $P-$a.s. two independent Brownian motions in $H$ with
the same covariance. We call $P$ a two sided Wiener probability
measure and $\left( W\left( t\right)  \right) _{t\in\mathbb{R}}$ a
two sided Brownian motion. Such objects exist, for every given
covariance operator, and play an important role in the theory of
random dynamical systems, see \cite{Arnold}. Details on
infinite-dimensional Brownian motions and their stochastic
integration can be found in \cite{DaPZ}. We will also denote by $E$
the expectation on $\left( \Omega ,\mathcal{F},P\right)  $.

For simplicity of the computations, and in analogy with equation
(\ref{GOY}), we assume that the components $\left(  W_{n}\left(
t\right)\right) _{t\in\mathbb{R}}$, for all $n \ge 1$, of the
two-sided Brownian motion have the form
\[
W_{n}\left(  t\right)  =\sigma_{n}\beta_{n}\left(  t\right)
\]
where $\beta_{n}\left(  t\right)  $ are independent two-sided
complex Brownian motions on $\left(  \Omega,\mathcal{F},P\right)  $
(with incremental covariance equal to one) and $\left(
\sigma_{n}\right) $ is the sequence given above.

On the probability space $\left(  \Omega,\mathcal{F},P\right)  $
consider the family of transformations $\left\{
\theta_{t}:\Omega\longmapsto\Omega,\ t\in \mathbb{R}\right\}  $
defined as $\theta_{t}\omega=\omega\left( t+\cdot\right)
-\omega\left(  t\right)  $, for every $\omega \in \Omega$. They are
measure preserving and ergodic with respect to $P$, and satisfy
$\theta_{0}=Id$, $\theta_{t+s}=\theta_{t}\circ\theta _{s},$ for
$s,t\in\mathbb{R}$, see \cite{Arnold}.

Let $\left(  \mathcal{F}_{t}\right)  _{t\in\mathbb{R}}$ be the
filtration associated to $\left(  W\left(  t\right)  \right)
_{t\in\mathbb{R}}$ ($\mathcal{F}_{t}$ is generated by $W\left(
s\right)  $ for all $s\leq t$).

Given initial conditions $u_{0},w_{0}\in H$, let us first rewrite system
(\ref{final_system}) in the abstract form%

\begin{eqnarray}\label{final_system_bis}\left\{
\begin{array}
[c]{ll}%
du^{\lambda}=[-\nu Au^{\lambda}-B(u^{\lambda},u^{\lambda})-\lambda
B(w^{\lambda},u^{\lambda})]dt+dW & \\
dw^{\lambda}=[-\nu Aw^{\lambda}-B(u^{\lambda},w^{\lambda})-\lambda
B(w^{\lambda},w^{\lambda})]dt+dW & \\
u^{\lambda}(0)=u_{0} & \\
w^{\lambda}(0)=w_{0} &
\end{array}\,.
\right. \end{eqnarray}

We consider the above Cauchy problem on $[0,\infty)$. Using the
notation of
the previous section, system (\ref{final_system_bis}) can be rewritten as follows%

\begin{align}
d\widetilde{u}^{\lambda}+\left(  \nu\widetilde{A}\widetilde{u}^{\lambda
}+\widetilde{B}_{\lambda}\left(  \widetilde{u}^{\lambda},\widetilde
{u}^{\lambda}\right)  \right)  dt &  =d\widetilde{W}\label{GOY_lambda}\\
\widetilde{u}^{\lambda}(0) &  =\widetilde{u}_{0}\nonumber
\end{align}
where $\widetilde{u}^{\lambda}\left(  t\right) =(u^{\lambda}\left(
t\right) ,w^{\lambda}\left(  t\right)  )$,
$\widetilde{u}_{0}=(u_{0},w_{0})$, $\widetilde{W}\left(  t\right)
=(W\left(  t\right)  ,W\left(  t\right)  )$. As an introductory
step, let us first give the usual definition of solution of
\eqref{GOY_lambda}; however, we will eventually need a more refined
notion of solution, that we will introduce  in Definition \ref{def
flow} below.

\begin{definition}
\label{defi weak solution u}Given
$\widetilde{u}_{0}\in\widetilde{H}$, we say that a stochastic
process $\widetilde{u}^{\lambda}(t,\omega)$ is a solution of
equation (\ref{GOY_lambda}) if it is a continuous adapted process in
$\widetilde{H}$ on $\left(  \Omega,\mathcal{F},\left(
\mathcal{F}_{t}\right) _{t\geq0},P\right) $ and, for $P$-a.e.
$\omega\in\Omega$,
\[
\widetilde{u}^{\lambda}(\cdot,\omega)\in C([0,T];\widetilde{H})\cap
L^{2}(0,T;\widetilde{V})\quad\text{for all }T>0
\]%
\begin{align*}
\left\langle \widetilde{u}^{\lambda}(t,\omega),\psi\right\rangle
_{\widetilde{H}} &  +\int_{0}^{t}\nu\left\langle
\widetilde{u}^{\lambda }(s,\omega),\widetilde{A}\psi\right\rangle
_{\widetilde{V},\widetilde
{V}^{\prime}}ds\\
&  +\int_{0}^{t}\left\langle \widetilde{B}_{\lambda}\left(  \widetilde
{u}^{\lambda}(s,\omega),\widetilde{u}^{\lambda}(s,\omega)\right)
,\psi\right\rangle _{\widetilde{H}}ds\\
&  =\left\langle \widetilde{u}_{0},\psi\right\rangle _{\widetilde{H}%
}+\left\langle \widetilde{W}(t,\omega),\psi\right\rangle _{\widetilde{H}}%
\end{align*}
for $t\geq0$ and $\psi\in\widetilde{V}$.
\end{definition}

Notice that $\widetilde{u}^{\lambda}(s,\omega)\in\widetilde{V}$ for a.e.
$s\geq0$, hence the integral of the bilinear term is well defined.

The above  definition is sufficient to analyze individual solutions,
but the theory of random attractors requires the concept of
stochastic flow: the $P$-negligible set where the properties of the
above definition may not hold, that is, it must be independent of
$\widetilde{u}_{0}$, and for $P$-a.e. $\omega \in\Omega$, moreover,
we will also need continuity with respect to the intial value
$\widetilde{u}_{0}$. In addition, here, we want to ``vary'' the
parameter $\lambda$ independently of $\omega$: {\it a priori} this
is not possible, again because the $P$-negligible set where the
properties of the previous definition hold, may depend on $\lambda$.
Both problems can be solved because it is possible to perform a
complete pathwise analysis of the equation. Let us, therefore, give
a more appropriate definition of solution for \eqref{GOY_lambda},
which is relevant to the above mentioned issues.

\begin{definition}
\label{def flow} A stochastic flow depending on $\lambda\in
\mathbb{R}$, associated with  equation (\ref{GOY_lambda}), is a
family of mappings $\left\{  \varphi
^{\lambda}(t,\omega):\widetilde{H}\rightarrow\widetilde{H};t\geq0,\omega
\in\Omega^{0},\lambda\in
\mathbb{R}\right\}  $, where $\Omega^{0}\in \mathcal{F}$ is $\theta_{t}%
$-invariant and $P\left(  \Omega^{0}\right)  =1$, with the
properties:

\begin{enumerate}
\item  for every $\lambda\in
\mathbb{R}$ and $\widetilde{u}_{0}\in\widetilde{H}$, $\left(
t,\omega\right) \mapsto\varphi^{\lambda}(t,\omega)\widetilde{u}_{0}$
(arbitrarily extended to all $\omega\in\Omega$) is a continuous
adapted process in $\widetilde{H}$ on $\left(
\Omega,\mathcal{F},\left( \mathcal{F}_{t}\right) _{t\geq0},P\right)
$ and for every $\omega\in\Omega^{0}$ we have
\[
\varphi^{\lambda}(\cdot,\omega)\widetilde{u}_{0}\in
C([0,T];\widetilde{H})\cap L^{2}(0,T;\widetilde{V}),\quad\text{for
all }T>0,
\]%
and
\begin{align*}
&\left\langle \varphi^{\lambda}(t,\omega)\widetilde{u}_{0},\psi
\right\rangle _{\widetilde{H}}+\int_{0}^{t}\nu\left\langle
\varphi^{\lambda
}(s,\omega)\widetilde{u}_{0},\widetilde{A}\psi\right\rangle
_{\widetilde {V},\widetilde{V}^{\prime}}ds\\
&+\int_{0}^{t}\left\langle \widetilde{B}_{\lambda}\left(
\varphi^{\lambda
}(s,\omega)\widetilde{u}_{0},\varphi^{\lambda}(s,\omega)\widetilde{u}%
_{0}\right)  ,\psi\right\rangle _{\widetilde{H}}ds
 =\left\langle \widetilde{u}_{0},\psi\right\rangle _{\widetilde{H}%
}+\left\langle \widetilde{W}(t,\omega),\psi\right\rangle _{\widetilde{H}}%
\end{align*}
for $t\geq0$ and $\psi\in\widetilde{V}$;

\item  for every $\lambda\in
\mathbb{R}$ and $\omega\in\Omega^{0}$, $\varphi^{\lambda
}(t,\omega)$ is a continuous map from  $\widetilde{H}$ into itself,
for all $t\geq0$; and
\[
\varphi^{\lambda}(t+s,\omega)=\varphi^{\lambda}(t,\theta_{s}\omega
)\circ\varphi^{\lambda}(s,\omega)
\]
for all $t,s\geq0$.
\end{enumerate}
\end{definition}

To emphasize the role of $\Omega^{0}$ in Definition \ref{def flow},
we will consider stochastic flows depending on $\lambda\in
\mathbb{R}$, \textit{defined on the set} $\Omega^{0}$.

We have the following result. The concept of uniqueness of
stochastic flow depending on $\lambda$ means:\ if we have two
stochastic flows, defined on two sets $\Omega_{1}^{0}$ and
$\Omega_{2}^{0}$, then they coincide on a set $\Omega_{3}^{0}\in
\mathcal{F}$ such that $P\left(  \Omega_{3}^{0}\right)  =1$.

\begin{theorem}
\label{theorem_goy_existence} Under assumption (\ref{assumption on
Q}), there exists a unique stochastic flow depending on $\lambda\in
\mathbb{R}$, in the sense of Defintion \ref{def flow}, associated
with equation (\ref{GOY_lambda}).
\end{theorem}

\begin{proof}
\textbf{Step 1} (preliminary facts). Denote by
$e^{-\nu\widetilde{A}t}$ the analytic semigroup generated by
$\widetilde{A}$ (see, e.g., \cite{Pazy}). By the general theory of
analytic semigroups or by explicit computation based on the spectral
representation, for every $\alpha>0$ we have $\left|
\widetilde{A}^{\alpha
}e^{-\nu\widetilde{A}t}\right|  _{\widetilde{H}}\leq\frac{C_{\alpha}%
}{t^{\alpha}}$ for some constant $C_{\alpha}>0$. Moreover, notice that
$\left\|  x\right\|  _{\widetilde{V}}=\left\|  \widetilde{A}^{1/2}x\right\|
_{\widetilde{H}}$ for all $x\in\widetilde{V}$.

The process $\widetilde{A}^{\varepsilon/2}\widetilde{W}\left(
t\right)  $ has $H$-components $\left(  A^{\varepsilon/2}W\left(
t\right)  ,A^{\varepsilon /2}W\left(  t\right)  \right)  $ where
$A^{\varepsilon/2}W\left(  t\right)  $ has complex components
$k_{n}^{\varepsilon}\sigma_{n}\beta_{n}\left( t\right)  $. Thanks to
assumption (\ref{assumption on Q}) it follows that
$\widetilde{A}^{\varepsilon/2}\widetilde{W}\left(  t\right)  $ is an
$\widetilde{H}$-valued Brownian motion. Thus it is
$\gamma$-H\"{o}lder continuous, with respect to $t$, in
$\widetilde{H}$, for every exponent $\gamma<\frac{1}{2}$, see
\cite{DaPZ}. This means that there exists a set $\Omega_{W}^{0}\in
\mathcal{F}$ such that $P\left( \Omega_{W}^{0}\right)  =1$ and
$\widetilde{A}^{\varepsilon /2}\widetilde{W}\left( t,\omega\right) $
is $\gamma$-H\"{o}lder continuous for every exponent
$\gamma<\frac{1}{2}$, for every $\omega\in\Omega_{W}^{0}$. The set
$\Omega_{W}^{0}$ is $\theta_{t}$-invariant, because H\"{o}lder
continuity is preserved by translation.

\textbf{Step 2} (auxiliary Stokes type problem). The pathwise
analysis of equation (\ref{GOY_lambda}) requires a careful
analysis of an auxiliary process. The process we are going to
introduce is usually defined as
\[
\widetilde{z}(t)=\int_{0}^{t}e^{-\nu\widetilde{A}(t-s)}d\widetilde{W}\left(
s\right)\, ,
\]
but from this definition via a stochastic integral (which is a
$P$-equivalence class) it is less easy to justify the
$\theta_{t}$-invariance of certain properties, on a full measure set
$\Omega^{0}$. For this reason we adopt the following less intuitive
definition. See \cite{Fland rendiconti} for further details on this
approach.

Let $\omega\in\Omega_{W}^{0}$ be given throughout this step, where $\Omega
_{W}^{0}$ has been defined in step 1. The function $t\mapsto\widetilde
{z}(t,\omega)$ given by
\begin{equation}\label{z-tilde}
\widetilde{z}(t,\omega)=e^{-\nu\widetilde{A}t}\widetilde{W}\left(
t,\omega\right)  +\int_{0}^{t}\nu\widetilde{A}e^{-\nu\widetilde{A}%
(t-s)}\left(  \widetilde{W}\left(  t,\omega\right)
-\widetilde{W}\left( s,\omega\right)  \right)  ds ,
\end{equation}
is well defined and bounded in $\widetilde{V}$, because (for
$\varepsilon$ that is given in assumption (\ref{assumption on Q}))
we have
\begin{align*}
\left\|  e^{-\nu\widetilde{A}t}\widetilde{W}\left(  t,\omega\right)  \right\|
_{\widetilde{V}}  & =\left|  \widetilde{A}^{1/2-\varepsilon/2}e^{-\nu
\widetilde{A}t}\widetilde{A}^{\varepsilon/2}\left(  \widetilde{W}\left(
t,\omega\right)  -\widetilde{W}\left(  0,\omega\right)  \right)  \right|
_{\widetilde{H}}\\
& \leq\left|  \widetilde{A}^{1/2-\varepsilon/2}e^{-\nu\widetilde{A}t}\right|
_{\widetilde{H}}\left|  \widetilde{A}^{\varepsilon/2}\left(  \widetilde
{W}\left(  t,\omega\right)  -\widetilde{W}\left(  0,\omega\right)  \right)
\right|  _{\widetilde{H}}\\
& \leq C\frac{1}{t^{1/2-\varepsilon/2}}t^{\beta}\, ,%
\end{align*}
for every $\beta<\frac{1}{2}$ and a suitable constant $C>0$ that
depends on $\beta$ and $\omega$. Observe that in the last estimate
we used the details described in step 1 above, in particular, the
H\"older continuity. Similarly
\begin{align*}
& \left\|  \int_{0}^{t}\widetilde{A}e^{-\nu\widetilde{A}(t-s)}\left(
\widetilde{W}\left(  t,\omega\right)  -\widetilde{W}\left(  s,\omega\right)
\right)  ds\right\|  _{\widetilde{V}}\\
& \leq\int_{0}^{t}\left|  \widetilde{A}^{3/2-\varepsilon/2}e^{-\nu
\widetilde{A}(t-s)}\widetilde{A}^{\varepsilon/2}\left(  \widetilde{W}\left(
t,\omega\right)  -\widetilde{W}\left(  s,\omega\right)  \right)  \right|
_{\widetilde{H}}ds\\
& \leq\int_{0}^{t}C\frac{1}{\left(  t-s\right)  ^{3/2-\varepsilon/2}}\left(
t-s\right)  ^{\beta}ds.
\end{align*}
The above estimates imply that
$\|\widetilde{z}(t,\omega)\|_{\widetilde{V}}$ is bounded on the
interval $[0,T]$, for all $T>0$ given, and the bound depends on $T$
and $\omega$.

With some additional minor effort one can show that the map
$t\mapsto\widetilde {z}(t,\omega)$ is continuous in $\widetilde{V}$.
We may write $\widetilde {z}(t,\omega)$ componentwise:
$\widetilde{z}(t,\omega)=\left( z^{\left( 1\right)
}(t,\omega),z^{\left(  2\right) }(t,\omega)\right)  $ where, in the
case when $\sigma_{n}\neq0$,
\[
z_{n}^{\left(  i\right)  }(t,\omega)=e^{-\nu k_{n}^{2}t}\sigma_{n}\beta
_{n}\left(  t,\omega\right)  +\int_{0}^{t}\nu k_{n}^{2}e^{-\nu k_{n}^{2}%
(t-s)}\left(  \sigma_{n}\beta_{n}\left(  t,\omega\right)  -\sigma_{n}\beta
_{n}\left(  s,\omega\right)  \right)  ds
\]
with $\beta_{n}\left(  t,\omega\right)  $ defined as $\sigma_{n}^{-1}%
W_{n}\left(  t,\omega\right)  $ (if $\sigma_{n}=0$, then
$z_{n}^{\left( i\right)  }=0$). From the componentwise identity it
is  easy  to deduce that
\begin{equation}
\left\langle \widetilde{z}(t,\omega),\psi\right\rangle _{\widetilde{H}%
}+\int_{0}^{t}\nu\left\langle \widetilde{z}(s,\omega),\widetilde{A}%
\psi\right\rangle
_{\widetilde{V},\widetilde{V}^{\prime}}ds=\left\langle
\widetilde{W}(t,\omega),\psi\right\rangle _{\widetilde{H}}\label{eq for z}%
\end{equation}
for all $t\geq0$ and $\psi\in\widetilde{V}$.

\textbf{Step 3} (auxiliary Navier-Stokes type random equation). Let
$\omega \in\Omega_{W}^{0}$ be given, and let
$\widetilde{z}(t,\omega))$  satisfy~(\ref{z-tilde}) or~(\ref{eq for
z}). Let us introduce the auxiliary random differential equation
\begin{align}\label{GOY_v_lambda}
\frac{d\widetilde{v}^{\lambda}(t,\omega)}{dt}+\nu\widetilde{A}\widetilde
{v}^{\lambda}(t,\omega)+\widetilde{B}_{\lambda}(\widetilde{v}^{\lambda
}(t,\omega)+\widetilde{z}(t,\omega),\widetilde{v}^{\lambda}(t,\omega
)+\widetilde{z}(t,\omega)) &  =0\\
\widetilde{v}^{\lambda}(0,\omega) &  =\widetilde{u}_{0} , \nonumber%
\end{align}
for $t\geq0$. We say that $\widetilde{v}^{\lambda}(\cdot,\omega)$ is
a weak solution of \eqref{GOY_v_lambda} if it belongs to
$C([0,T];\widetilde{H})\cap L^{2}(0,T;\widetilde {V})$, for all
$T>0$, and if in addition it satisfies
\[
\left\langle \widetilde{v}^{\lambda}(t,\omega),\psi\right\rangle
_{\widetilde{H}}+\int_{0}^{t}\nu\left\langle \widetilde{v}^{\lambda}%
(s,\omega),\widetilde{A}\psi\right\rangle
_{\widetilde{V},\widetilde {V}^{\prime}}ds
\]%
\[
+\int_{0}^{t}\left\langle \widetilde{B}_{\lambda}\left( \widetilde
{v}^{\lambda}(s,\omega)+\widetilde{z}(s,\omega),\widetilde{v}^{\lambda
}(s,\omega)+\widetilde{z}(s,\omega)\right)  ,\psi\right\rangle
_{\widetilde{H}}ds=\left\langle
\widetilde{u}_{0},\psi\right\rangle
_{\widetilde{H}} ,%
\]
for every $t\geq0$ and $\psi\in\widetilde{V}$.

For every $\omega\in\Omega_{W}^{0}$ and $\lambda\in \mathbb{R}$,
there exists a unique weak solution
$\widetilde{v}^{\lambda}(\cdot,\omega)=\widetilde{v}^{\lambda
}(\cdot,\omega,\widetilde{u}_{0})$ of  equation~(\ref{GOY_v_lambda})
and it depends continuously, in \\$C([0,T];\widetilde{H})\cap
L^{2}(0,T;\widetilde{V})$ norms, for any given $T>0$, on the initial
condition $\widetilde{u}_{0}$ in $\widetilde{H}$. A full rigorous
proof of this statement is very long, but at the same time it is
very classical. Similar detailed proofs are given, for instance, in
\cite{flandoli-goy}, \cite{Fladissipative},  and in
\cite{Constantin-Foias} or \cite{Temam} in the case of the classical
Navier-Stokes equations (i.e., when $\widetilde{z}=0$). The rigorous
detailed proof is based on the Galerkin approximation procedure and
then passing to the limit using the appropriate compactness
theorems. We omit these details which can be found in the above
references. Instead, we present here
%
%
the formal computations which lead to the basic {\it a priori}
estimates, this is in order to stress the role played by
$\widetilde{z}$. Formally, if
$\widetilde{v}^{\lambda}=\widetilde{v}^{\lambda}(t,\omega)$ is a
solution, then from various estimates and properties stated in Lemma
\ref{lemma su B tilde} we have
\begin{align*}
\frac{1}{2}\frac{d}{dt}\left|  \widetilde{v}^{\lambda}\right|  _{\widetilde
{H}}^{2}{}+\nu\left\|  \widetilde{v}^{\lambda}\right\|  _{\widetilde{V}}%
^{2}{}  & \leq|<\widetilde{B}_{\lambda}(\widetilde{v}^{\lambda}+\widetilde
{z},\widetilde{z}),\widetilde{v}^{\lambda}>_{\widetilde{H}}|\\
& \leq C\left|  \widetilde{v}^{\lambda}\right|  _{\widetilde{H}}\left\|
\widetilde{z}\right\|  _{\widetilde{V}}\left(  \left|  \widetilde{v}^{\lambda
}\right|  _{\widetilde{H}}+\left|  \widetilde{z}\right|  _{\widetilde{H}%
}\right)  .
\end{align*}
On a given interval $\left[  0,T\right]  $, $\left\|
\widetilde{z}\left( \cdot,\omega\right)  \right\| _{\widetilde{V}}$
and $\left|  \widetilde {z}\left( \cdot,\omega\right)  \right|
_{\widetilde{H}}$ are bounded (see step 2 above; they are bounded by
a constant depending on $\omega$), hence there is $C\left(
\omega\right)  >0$ such that
\[
\frac{1}{2}\frac{d}{dt}\left|  \widetilde{v}^{\lambda}\right|  _{\widetilde
{H}}^{2}{}+\nu\left\|  \widetilde{v}^{\lambda}\right\|  _{\widetilde{V}}^{2}%
{}\leq C\left(  \omega\right)  \left|  \widetilde{v}^{\lambda}\right|
_{\widetilde{H}}\left(  \left|  \widetilde{v}^{\lambda}\right|  _{\widetilde
{H}}+C\left(  \omega\right)  \right)
\]
which implies, by Gronwall lemma, a bound in terms of $C\left(  \omega\right)
$ and $\left|  \widetilde{u}_{0}\right|  _{\widetilde{H}}^{2}$ for $\sup
_{t\in\left[  0,T\right]  }\left|  \widetilde{v}^{\lambda}\left(
t,\omega\right)  \right|  _{\widetilde{H}}^{2}$ and $\int_{0}^{T}\left\|
\widetilde{v}^{\lambda}\left(  t,\omega\right)  \right\|  _{\widetilde{V}}%
^{2}dt$. These are the basic {\it a priori} bounds for the existence
of weak solutions. For uniqueness and continuous dependence on
initial data, we consider $\widetilde{v}_{1}^{\lambda}\left( \cdot
,\omega\right)  $ and $\widetilde{v}_{2}^{\lambda}\left(
\cdot,\omega\right) $ two weak solutions, and we set
$\widetilde{y}^{\lambda}\left( t,\omega\right)
=\widetilde{v}_{1}^{\lambda}\left( t,\omega\right)
-\widetilde{v}_{2}^{\lambda}\left( t,\omega\right) $, then formally
we have
\[
\frac{d\widetilde{y}^{\lambda}}{dt}+\nu\widetilde{A}\widetilde{y}^{\lambda
}+\widetilde{B}_{\lambda}\left(  \widetilde{y}^{\lambda},\widetilde{v}%
_{1}^{\lambda}+\widetilde{z}\right)  +\widetilde{B}_{\lambda}\left(
\widetilde{v}_{2}^{\lambda}+\widetilde{z},\widetilde{y}^{\lambda}\right)\,
.
\]
Thus, by Lemma \ref{lemma su B tilde} and the boundedness of $\left|
\widetilde{z}\left(  \cdot,\omega\right)  \right| _{\widetilde{H}}$
and $\left|  \widetilde{v}_{1}^{\lambda}\left( \cdot,\omega\right)
\right| _{\widetilde{H}}$ on a given $\left[ 0,T\right]  $,  we
formally have
\begin{align*}
\frac{1}{2}\frac{d}{dt}\left|  \widetilde{y}^{\lambda}\right|  _{\widetilde
{H}}^{2}{}+\nu\left\|  \widetilde{y}^{\lambda}\right\|  _{\widetilde{V}}%
^{2}{}  & \leq|<\widetilde{B}_{\lambda}(\widetilde{y}^{\lambda},\widetilde
{v}_{1}^{\lambda}+\widetilde{z}),\widetilde{y}^{\lambda}>_{\widetilde{H}}|\\
& \leq C\left|  \widetilde{y}^{\lambda}\right|  _{\widetilde{H}}\left\|
\widetilde{y}^{\lambda}\right\|  _{\widetilde{V}}\left(  \left|  \widetilde
{v}_{1}^{\lambda}\right|  _{\widetilde{H}}+\left|  \widetilde{z}\right|
_{\widetilde{H}}\right)  \\
& \leq\frac{\nu}{2}\left\|  \widetilde{y}^{\lambda}\right\|  _{\widetilde{V}%
}^{2}+\frac{1}{\nu}C\left(  \nu,\omega\right)  \left|  \widetilde{y}^{\lambda
}\right|  _{\widetilde{H}}^{2}%
\end{align*}
for some constant $C( \nu, \omega)  >0$. Again by Gronwall Lemma,
uniqueness and continuous dependence on initial value follow.

\textbf{Step 4} (existence of the stochastic flow). For every
$\lambda\in \mathbb{R}$, $t\geq0$, $\omega\in\Omega_{W}^{0}$ and
$\widetilde{u}_{0}\in\widetilde{H}$, define
\[
\varphi^{\lambda}(t,\omega)\widetilde{u}_{0}=\widetilde{v}^{\lambda}\left(
t,\omega,\widetilde{u}_{0}\right)  +\widetilde{z}\left(  t,\omega\right)
\]
where $\widetilde{v}^{\lambda}\left(
\cdot,\omega,\widetilde{u}_{0}\right)  $ is the unique weak solution
given in step 3 and $\widetilde{z}\left(  \cdot ,\omega\right)  $ is
defined in step 2. The set $\Omega_{W}^{0}$ is
$\theta_{t}$-invariant and $P\left(  \Omega_{W}^{0}\right) =1$.
Property 1 of Definition \ref{def flow} is a direct consequence of
the analogous properties of $\widetilde{v}^{\lambda}\left(
\cdot,\omega,\widetilde{u}_{0}\right)  $ and $\widetilde{z}\left(
\cdot,\omega\right)  $ proved in steps 2 and 3. As to property 2 of
Definition \ref{def flow}, given $\lambda\in \mathbb{R}$, $\omega
\in\Omega_{W}^{0}$, $t\geq0$, the continuity of
$\varphi^{\lambda}(t,\omega)$ in $\widetilde{H}$ is a consequence of
the continuous dependence of $\widetilde{v}^{\lambda}\left(
\cdot,\omega,\widetilde{u}_{0}\right)  $ on $\widetilde{u}_{0}$, see
step 3. The property
\begin{equation}
\varphi^{\lambda}(t+s,\omega)\widetilde{u}_{0}=\varphi^{\lambda}(t,\theta
_{s}\omega)\varphi^{\lambda}(s,\omega)\widetilde{u}_{0}\label{flow property}%
\end{equation}
for all $t,s\geq0$ follows from the uniqueness statement of step 3.
In order to prove this claim, let us write, for a given $s\geq0$,
the equation satisfied by the two functions
$t\mapsto\varphi^{\lambda}(t+s,\omega)\widetilde{u}_{0}$ and
$t\mapsto\varphi^{\lambda}(t,\theta_{s}\omega)\varphi^{\lambda}(s,\omega)\widetilde{u}_{0}$
for $t\geq0$. We know that
$\varphi^{\lambda}(t,\omega)\widetilde{u}_{0}$ satisfies the weak
form of the equation given in Definition \ref{def flow}. From it,
for the function $y\left(  t\right)  :=\varphi^{\lambda}(t+s,\omega
)\widetilde{u}_{0}$, we have
\begin{align*}
& \left\langle y\left(  t\right)  ,\psi\right\rangle _{\widetilde{H}}%
+\int_{0}^{t+s}\nu\left\langle \varphi^{\lambda}(r,\omega)\widetilde{u}%
_{0},\widetilde{A}\psi\right\rangle _{\widetilde{V},\widetilde{V}^{\prime}%
}dr\\
& +\int_{0}^{t+s}\left\langle \widetilde{B}_{\lambda}\left(  \varphi^{\lambda
}(r,\omega)\widetilde{u}_{0},\varphi^{\lambda}(r,\omega)\widetilde{u}%
_{0}\right)  ,\psi\right\rangle _{\widetilde{H}}dr\\
& =\left\langle \widetilde{u}_{0},\psi\right\rangle _{\widetilde{H}%
}+\left\langle \widetilde{W}(t+s,\omega),\psi\right\rangle _{\widetilde{H}}%
\end{align*}
for all $t\geq0$ and $\psi\in\widetilde{V}$. Hence,
\begin{align*}
& \left\langle y\left(  t\right)  ,\psi\right\rangle _{\widetilde{H}}%
+\int_{0}^{t}\nu\left\langle y\left(  r\right) ,\widetilde{A}\psi
\right\rangle _{\widetilde{V},\widetilde{V}^{\prime}}dr+\int_{0}%
^{t}\left\langle \widetilde{B}_{\lambda}\left(  y\left(  r\right)  ,y\left(
r\right)  \right)  ,\psi\right\rangle _{\widetilde{H}}dr\\
& =\left\langle \widetilde{u}_{0},\psi\right\rangle _{\widetilde{H}%
}+\left\langle \widetilde{W}(t+s,\omega),\psi\right\rangle _{\widetilde{H}%
}+\left\langle
\varphi^{\lambda}(s,\omega)\widetilde{u}_{0},\psi\right\rangle
_{\widetilde{H}}\\
& -\left\langle \widetilde{u}_{0},\psi\right\rangle _{\widetilde{H}%
}-\left\langle \widetilde{W}(s,\omega),\psi\right\rangle
_{\widetilde{H}}\, .%
\end{align*}
That is,
\begin{align*}
& \left\langle y\left(  t\right)  ,\psi\right\rangle _{\widetilde{H}}%
+\int_{0}^{t}\nu\left\langle y\left(  r\right) ,\widetilde{A}\psi
\right\rangle _{\widetilde{V},\widetilde{V}^{\prime}}dr+\int_{0}%
^{t}\left\langle \widetilde{B}_{\lambda}\left(  y\left(  r\right)  ,y\left(
r\right)  \right)  ,\psi\right\rangle _{\widetilde{H}}dr\\
& =\left\langle
\varphi^{\lambda}(s,\omega)\widetilde{u}_{0},\psi\right\rangle
_{\widetilde{H}}+\left\langle \left[
\widetilde{W}(t+s,\omega)-\widetilde {W}(s,\omega)\right]
,\psi\right\rangle _{\widetilde{H}}.
\end{align*}
Recall that $\widetilde{W}\left(  t,\omega\right)  =\left(  W\left(
t,\omega\right)  ,W\left(  t,\omega\right)  \right)  $, $W\left(
t,\omega\right)  =\omega\left(  t\right)  $, and
\begin{align*}
W\left(  t,\theta_{s}\omega\right)    & =\theta_{s}\omega\left(  t\right)
=\omega\left(  s+t\right)  -\omega\left(  s\right)  \\
& =W\left(  s+t,\omega\right)  -W\left(  s,\omega\right)  .
\end{align*}
Therefore, $y$ satisfies
\begin{align*}
& \left\langle y\left(  t\right)  ,\psi\right\rangle _{\widetilde{H}}%
+\int_{0}^{t}\nu\left\langle y\left(  r\right) ,\widetilde{A}\psi
\right\rangle _{\widetilde{V},\widetilde{V}^{\prime}}dr+\int_{0}%
^{t}\left\langle \widetilde{B}_{\lambda}\left(  y\left(  r\right)  ,y\left(
r\right)  \right)  ,\psi\right\rangle _{\widetilde{H}}dr\\
& =\left\langle
\varphi^{\lambda}(s,\omega)\widetilde{u}_{0},\psi\right\rangle
_{\widetilde{H}}+\left\langle \widetilde{W}(t,\theta_{s}\omega),\psi
\right\rangle _{\widetilde{H}}.
\end{align*}
This is the same equation satisfied by the map $t\mapsto\varphi^{\lambda}(t,\theta_{s}%
\omega)\varphi(s,\omega)\widetilde{u}_{0}$ for $t\geq0$. Since this
equation corresponds to the auxiliary equation of step 3 through the
transformation via $\widetilde{z}\left( t,\omega\right)  $ (the
detailed argument is the same as
the one given below in step 5, but easier since here $\Omega_{1}^{0}%
=\Omega_{W}^{0}$, see below) and we have uniqueness for the
latter, thus we also have uniqueness for the former. This proves
(\ref{flow property}).

\textbf{Step 5} (uniqueness of the stochastic flow). Let $\varphi_{1}%
^{\lambda}(t,\omega)$ be a stochastic flow depending on $\lambda\in
\mathbb{R}$, associated with equation (\ref{GOY_lambda}), defined on
a $\theta_{t}$-invariant full measure set $\Omega_{1}^{0}$. Consider
the $\theta_{t}$-invariant full measure set $\Omega_{W}^{0}$
described in step 2 above. The set $\Omega
_{1}^{0}\cap\Omega_{W}^{0}$ is $\theta_{t}$-invariant and $P\left(
\Omega
_{1}^{0}\cap\Omega_{W}^{0}\right)  =1$. For all $\omega\in\Omega_{1}^{0}%
\cap\Omega_{W}^{0}$ define
\[
\widetilde{v}_{1}^{\lambda}\left(  t,\omega,\widetilde{u}_{0}\right)
:=\varphi_{1}^{\lambda}(t,\omega)\widetilde{u}_{0}-\widetilde{z}\left(
t,\omega\right)  .
\]
From the properties of $\varphi_{1}^{\lambda}$ and $\widetilde{z}$ it is
trivial to check that $\widetilde{v}_{1}^{\lambda}\left(  \cdot,\omega
,\widetilde{u}_{0}\right)  $ is a weak solution of the auxiliary equation of
step 3. Of course we have
\[
\varphi_{1}^{\lambda}(t,\omega)\widetilde{u}_{0}=\widetilde{v}_{1}^{\lambda
}\left(  t,\omega,\widetilde{u}_{0}\right)  +\widetilde{z}\left(
t,\omega\right)  .
\]

Now, let $\varphi_{2}^{\lambda}(t,\omega)$ be another stochastic
flow depending on $\lambda\in \mathbb{R}$, associated with equation
(\ref{GOY_lambda}), with its $\theta_{t}$-invariant full measure set
$\Omega_{2}^{0}$. The function
\[
\widetilde{v}_{2}^{\lambda}\left(  t,\omega,\widetilde{u}_{0}\right)
:=\varphi_{2}^{\lambda}(t,\omega)\widetilde{u}_{0}-\widetilde{z}\left(
t,\omega\right)
\]
defined for $\omega\in\Omega_{2}^{0}\cap\Omega_{W}^{0}$ is a weak solution of
the auxiliary equation of step 3. Thus, for $\omega\in\Omega_{1}^{0}\cap
\Omega_{2}^{0}\cap\Omega_{W}^{0}$ we have
\[
\widetilde{v}_{1}^{\lambda}\left(  t,\omega,\widetilde{u}_{0}\right)
=\widetilde{v}_{2}^{\lambda}\left(  t,\omega,\widetilde{u}_{0}\right)
\]
because of the uniqueness of solutions for equation
\eqref{GOY_v_lambda}, for every $\omega \in\Omega_{W}^{0}$. Thus
\[
\varphi_{1}^{\lambda}(t,\omega)\widetilde{u}_{0}=\varphi_{2}^{\lambda
}(t,\omega)\widetilde{u}_{0}%
\]
for all $\omega\in\Omega_{1}^{0}\cap\Omega_{2}^{0}\cap\Omega_{W}^{0}$. Since
$P\left(  \Omega_{1}^{0}\cap\Omega_{2}^{0}\cap\Omega_{W}^{0}\right)  =1$, the
proof is complete.
\end{proof}

In step 5 of the previous proof we have obtained also the following
representation result, which will be useful in the next section.

\begin{corollary}
\label{corollary flow}Let $\varphi^{\lambda}(t,\omega)$ be a
stochastic flow depending on $\lambda\in \mathbb{R}$, associated
with equation (\ref{GOY_lambda}), defined on a
$\theta_{t}$-invariant full measure set $\Omega_{1}^{0}$. On the
$\theta_{t}$-invariant full measure set $\Omega_{W}^{0}$ described
in step 2 of Theorem \ref{theorem_goy_existence}, one can define the
functions $\widetilde{z}\left(  t,\omega\right)  $ and
$\widetilde{v}^{\lambda}\left( t,\omega,\widetilde{u}_{0}\right) $
according to steps 2 and 3 of that proof. Then, on the
$\theta_{t}$-invariant full measure set
$\Omega^{0}:=\Omega_{1}^{0}\cap \Omega_{W}^{0}$ we have
\[
\varphi^{\lambda}(t,\omega)\widetilde{u}_{0}=\widetilde{v}^{\lambda}\left(
t,\omega,\widetilde{u}_{0}\right)  +\widetilde{z}\left(  t,\omega\right)  .
\]
\end{corollary}

\subsection{Continuous dependence with respect to the parameter $\lambda$}

As above, we assume condition (\ref{assumption on Q}). The
uniformity in the initial condition of the next statement will be
used to prove the upper semicontinuity of the random attractor with
respect to the parameter $\lambda$.

\begin{theorem}
Let $\varphi^{\lambda}(t,\omega)$ be the stochastic flow that was
established in Theorem \ref{theorem_goy_existence} and Corollary
\ref{corollary flow},  associated with equation (\ref{GOY_lambda})
and depending on the parameter $\lambda\in \mathbb{R}$. Let
$\Omega^{0}\in \mathcal{F}$, $P\left( \Omega^{0}\right) =1$, be a
$\theta_{t}$-invariant set where all the properties of Definition
\ref{def flow} and Corollary \ref{corollary flow} hold true. Then,
for every $\omega\in\Omega^{0}$, we have
\[
\lim_{\lambda\longrightarrow\lambda_{0}}\sup_{\widetilde{u}_{0}\in B}%
\sup_{0\leq t\leq T}\left|  \varphi^{\lambda}(t,\omega)\widetilde{u}%
_{0}-\varphi^{\lambda_{0}}(t,\omega)\widetilde{u}_{0}\right|  _{\widetilde{H}%
}^{{}}=0
\]
for all $T>0$, $\lambda_{0}\in \mathbb{R}$ and all bounded sets
$B\subset\widetilde{H}$.
\end{theorem}

\begin{proof}
We prove the theorem only in the case $\lambda_{0}=0$, the general
case being the same. The elements $\omega\in\Omega^{0}$ and $T>0$
are given and fixed throughout the proof, as well as the bounded set
$B\subset\widetilde{H}$.

\textbf{Step 1 }(preparation). Denote by $\left(
u^{\lambda}(t,\omega
,\widetilde{u}_{0}),w^{\lambda}(t,\omega,\widetilde{u}_{0})\right) $
the decomposition of $\varphi^{\lambda}(t,\omega)\widetilde{u}_{0}$
in $\widetilde{H}=H\times H$, and by $\left(  u_{0},w_{0}\right)  $
the decomposition of the initial value $\widetilde{u}_{0}$. Where it
is necessary, we will shorten the notation and write $\left(
u^{\lambda}(t),w^{\lambda}(t)\right)  $ and apply analogous change
of notation to other similar quantities.

From the weak integral equation in Definition \ref{def flow} and
the definitions of $\widetilde{A}$ and $\widetilde{B}_{\lambda}$
we have
\begin{align*}
& \left\langle u^{\lambda}(t),\psi_{1}\right\rangle _{H}+\int_{0}^{t}%
\nu\left\langle u^{\lambda}(s),A\psi_{1}\right\rangle _{V,V^{\prime}}ds\\
& +\int_{0}^{t}\left\langle B\left(
u^{\lambda}(s),u^{\lambda}(s)\right) +\lambda B\left(
w^{\lambda}(s),u^{\lambda}(s)\right)  ,\psi_{1}\right\rangle
_{H}ds\\
& =\left\langle u_{0},\psi_{1}\right\rangle _{H}+\left\langle
W(t,\omega ),\psi_{1}\right\rangle _{H}\, ,%
\end{align*}
for all $t\geq0$ and $\psi_{1}\in V$ and
\begin{align*}
& \left\langle w^{\lambda}(t),\psi_{2}\right\rangle _{H}+\int_{0}^{t}%
\nu\left\langle w^{\lambda}(s),A\psi_{2}\right\rangle _{V,V^{\prime}}ds\\
& +\int_{0}^{t}\left\langle B\left(
u^{\lambda}(s),w^{\lambda}(s)\right) +\lambda B\left(
w^{\lambda}(s),w^{\lambda}(s)\right)  ,\psi_{2}\right\rangle
_{H}ds\\
& =\left\langle w_{0},\psi_{2}\right\rangle _{H}+\left\langle
W(t,\omega ),\psi_{2}\right\rangle _{H}\, ,%
\end{align*}
for all $t\geq0$ and $\psi_{2}\in V$. Let us define the new
function
\[
q^{\lambda}(t,\omega,\widetilde{u}_{0}):=u^{\lambda}(t,\omega,\widetilde
{u}_{0})+\lambda w^{\lambda}(t,\omega,\widetilde{u}_{0})
\]
and the corresponding difference
\[
\rho^{\lambda}(t,\omega,\widetilde{u}_{0}):=q^{\lambda}(t,\omega,\widetilde
{u}_{0})-q^{0}(t,\omega,\widetilde{u}_{0})=q^{\lambda}(t,\omega,\widetilde
{u}_{0})-u^{0}(t,\omega,\widetilde{u}_{0}).
\]
The above quantities  are solutions, respectively, of
\begin{align*}
& \left\langle q^{\lambda}(t),\psi_{1}\right\rangle _{H}+\int_{0}^{t}%
\nu\left\langle q^{\lambda}(s),A\psi_{1}\right\rangle _{V,V^{\prime}}ds\\
& +\int_{0}^{t}\left\langle B\left(
q^{\lambda}(s),q^{\lambda}(s)\right)
,\psi_{1}\right\rangle _{H}ds\\
& =\left\langle u_{0}+\lambda w_{0},\psi_{1}\right\rangle
_{H}+\left( 1+\lambda\right)  \left\langle
W(t,\omega),\psi_{1}\right\rangle _{H} \, ,%
\end{align*}
for $t\geq0$ and $\psi_{1}\in V$, and
\begin{align*}
& \left\langle \rho^{\lambda}(t),\psi_{2}\right\rangle _{H}+\int_{0}^{t}%
\nu\left\langle \rho^{\lambda}(s),A\psi_{2}\right\rangle _{V,V^{\prime}}ds\\
& +\int_{0}^{t}\left\langle B\left(
q^{\lambda}(s),\rho^{\lambda}(s)\right) +B\left(
\rho^{\lambda}(s),q^{\lambda}(s)\right)  -B\left(  \rho^{\lambda
}(s),\rho^{\lambda}(s)\right)  ,\psi_{2}\right\rangle _{H}ds\\
& =\left\langle \lambda w_{0},\psi_{2}\right\rangle
_{H}+\lambda\left\langle W(t,\omega),\psi_{2}\right\rangle _{H}\, ,%
\end{align*}
for $t\geq0$ and $\psi_{2}\in V$.

\textbf{Step 2 }(bound on $q^{\lambda}$). Let us prove next that
\[
\sup_{\lambda\in\left[  -1,1\right]  }\sup_{\widetilde{u}_{0}\in
B}\sup_{0\leq t\leq T}\left|
q^{\lambda}(t,\omega,\widetilde{u}_{0})\right|  _{H}<\infty.
\]
Define
\[
\overline{v}^{\lambda}(t,\omega,\widetilde{u}_{0}):=q^{\lambda}(t,\omega
,\widetilde{u}_{0})-\left(  1+\lambda\right)  z(t,\omega)\, ,
\]
where $z(t,\omega)$ is any one of the two equal components of
$\widetilde {z}(t,\omega)$ given in Corollary \ref{corollary
flow}. We have
\begin{align*}
& \left\langle \overline{v}^{\lambda}(t),\psi_{1}\right\rangle
_{H}+\int _{0}^{t}\nu\left\langle
\overline{v}^{\lambda}(s),A\psi_{1}\right\rangle
_{V,V^{\prime}}ds\\
& +\int_{0}^{t}\left\langle B\left(  \overline{v}^{\lambda}(s)+\left(
1+\lambda\right)  z(s),\overline{v}^{\lambda}(s)+\left(  1+\lambda\right)
z(s)\right)  ,\psi_{1}\right\rangle _{H}ds\\
& =\left\langle u_{0}+\lambda w_{0},\psi_{1}\right\rangle _{H}\, ,%
\end{align*}
for all $t\geq0$ and $\psi_{1}\in V$. Formally, this implies
\[
\frac{1}{2}\frac{d}{dt}\left|  \overline{v}^{\lambda}\right|  _{H}^{2}%
+\nu\left\|  \overline{v}^{\lambda}\right\|  _{V}^{2}\leq\left|
\left\langle B\left(  \overline{v}^{\lambda}+\left( 1+\lambda\right)
z,\overline {v}^{\lambda}+\left(  1+\lambda\right) z\right)
,\overline{v}^{\lambda }\right\rangle _{H}\right| \, ,
\]
and thus
\begin{align*}
& \frac{1}{2}\left|  \overline{v}^{\lambda}\left(  t\right)  \right|  _{H}%
^{2}+\nu\int_{0}^{t}\left\|  \overline{v}^{\lambda}\left(  s\right)  \right\|
_{V}^{2}ds\leq\frac{1}{2}\left|  u_{0}+\lambda w_{0}\right|  _{H}^{2}\\
& +\int_{0}^{t}\left|  \left\langle B\left(  \overline{v}^{\lambda}\left(
s\right)  +\left(  1+\lambda\right)  z\left(  s\right)  ,\overline{v}%
^{\lambda}\left(  s\right)  +\left(  1+\lambda\right)  z\left(  s\right)
\right)  ,\overline{v}^{\lambda}\left(  s\right)  \right\rangle _{H}\right|
ds.
\end{align*}
Rigorously, the above inequality can be proved either by general
abstract theorems (see \cite{Temam}) or by taking finite-dimensional
(i.e. with finite many components) test functions $\psi_{1}$,
performing the computations at the finite dimensional level and then
taking the limit, which can be justified because the map

$$s\mapsto \left\langle B\left( \overline{v}^{\lambda}\left(
s\right) +\left( 1+\lambda\right) z\left(  s\right)
,\overline{v}^{\lambda}\left( s\right) +\left( 1+\lambda\right)
z\left(  s\right)  \right) ,\overline{v}^{\lambda }\left( s\right)
\right\rangle _{H}$$ is integrable. Thus, from Lemma \ref{lemma su B
tilde} and the bounds on $z\left( t,\omega\right) $, given in step 2
of the proof of Theorem \ref{theorem_goy_existence}, we have, for
$t\in\left[ 0,T\right] $,
\begin{align*}
& \frac{1}{2}\left|  \overline{v}^{\lambda}\left(  t\right)  \right|  _{H}%
^{2}+\nu\int_{0}^{t}\left\|  \overline{v}^{\lambda}\left(  s\right)  \right\|
_{V}^{2}ds-\frac{1}{2}\left|  u_{0}+\lambda w_{0}\right|  _{H}^{2}\\
& \leq2\int_{0}^{t}\left|  \left\langle B\left(  \overline{v}^{\lambda}\left(
s\right)  +\left(  1+\lambda\right)  z\left(  s\right)  ,z\left(  s\right)
\right)  ,\overline{v}^{\lambda}\left(  s\right)  \right\rangle _{H}\right|
ds\\
& \leq C\int_{0}^{t}\left|  \overline{v}^{\lambda}\left(  s\right)  \right|
_{H}\left|  \overline{v}^{\lambda}\left(  s\right)  +\left(  1+\lambda\right)
z\left(  s\right)  \right|  _{H}\left\|  z\left(  s\right)  \right\|
_{V}ds\\
& \leq C\left(  \omega\right)  \int_{0}^{t}\left|
\overline{v}^{\lambda }\left(  s\right)  \right|  _{H}\left(  \left|
\overline{v}^{\lambda}\left( s\right)  \right|  _{H}+C\left(
\omega\right)  \right)  ds,
\end{align*}
where $C\left(  \omega\right)  $ depends on the bounds of the
relevant norms of  $z\left( \cdot,\omega\right) $ over the interval
$\left[ 0,T\right]$, which in principle is also depending on $T$. By
Gronwall lemma we deduce
\[
\left|  \overline{v}^{\lambda}\left(  t,\omega,\widetilde{u}_{0}\right)
\right|  _{H}\leq C\left(  \omega\right)  \cdot\left|  \overline{v}^{\lambda
}\left(  0,\omega,\widetilde{u}_{0}\right)  \right|  _{H}%
\]
on $\left[  0,T\right]$, for a new constant $C\left(  \omega\right)
$. This implies
\[
\left|  q^{\lambda}\left(  t,\omega,\widetilde{u}_{0}\right)
\right| _{H}\leq2\left|  z(t,\omega)\right|  _{H}+C\left(
\omega\right)  \cdot\left| u_{0}+\lambda w_{0}\right|  _{H}\, ,%
\]
on $\left[  0,T\right]  $, and the claim of this step is proved,
using again the bounds on $z\left(  \cdot,\omega\right)  $ over the
interval  $\left[ 0,T\right]  $.

\textbf{Step 3 }(convergence of $\rho^{\lambda}$). Next we prove
that
\[
\lim_{\lambda\longrightarrow0}\sup_{\widetilde{u}_{0}\in B}\sup_{0\leq t\leq
T}\left|  \rho^{\lambda}(t,\omega,\widetilde{u}_{0})\right|  _{H}^{{}}=0.
\]
In one sentence, this is a consequence of the various bounds, that
we have established previously,  and the fact that the initial
condition $\lambda w_{0}$ and the forcing term $\lambda W(t,\omega)$
converge to zero, as $\lambda\longrightarrow0$. Define
\[
\widehat{v}^{\lambda}(t,\omega,\widetilde{u}_{0}):=\rho^{\lambda}%
(t,\omega,\widetilde{u}_{0})-\lambda z(t,\omega),
\]
it satisfies
\begin{align*}
& \left\langle \widehat{v}^{\lambda}(t),\psi_{2}\right\rangle _{H}+\int_{0}%
^{t}\nu\left\langle
\widehat{v}^{\lambda}(s),A\psi_{2}\right\rangle
_{V,V^{\prime}}ds\\
& +\int_{0}^{t}\left\langle B\left(  q^{\lambda}(s),\rho^{\lambda}(s)\right)
+B\left(  \rho^{\lambda}(s),q^{\lambda}(s)\right)  -B\left(  \rho^{\lambda
}(s),\rho^{\lambda}(s)\right)  ,\psi_{2}\right\rangle _{H}ds\\
& =\left\langle \lambda w_{0},\psi_{2}\right\rangle _{H}\, ,%
\end{align*}
for all $t\geq0$ and $\psi_{2}\in V$. \ By virtue of Lemma
\ref{lemma estimate} and Lemma \ref{lemma su B tilde}, we have
\[
\left|  \left\langle B\left(  q^{\lambda},\rho^{\lambda}\right)
,\widehat {v}^{\lambda}\right\rangle _{H}\right|  =\left|
\left\langle B\left( q^{\lambda},\lambda z\right)
,\widehat{v}^{\lambda}\right\rangle _{H}\right| \leq
C|\lambda|\left\|  z\right\|  _{V}\left|  q^{\lambda}\right|
_{H}\left| \widehat{v}^{\lambda}\right|  _{H}\, .%
\]
Thus, using both the bound on $z\left(  \cdot,\omega\right)  $ on
$\left[ 0,T\right]  $ and the bound of the previous step, there is
constant $C_{1}\left(  \omega\right)  $ such that
\[
\left|  \left\langle B\left(  q^{\lambda},\rho^{\lambda}\right)
,\widehat {v}^{\lambda}\right\rangle _{H}\right|
\leq\lambda^{2}C_{1}\left( \omega\right)  +\left|
\widehat{v}^{\lambda}\right|  _{H}^{2}\, ,%
\]
for all $\widetilde{u}_{0}\in B$ and $t \in [0,T]$. Similarly,
\begin{align*}
\left|  \left\langle B\left(  \rho^{\lambda},q^{\lambda}\right)  ,\widehat
{v}^{\lambda}\right\rangle _{H}\right|    & =\left|  \left\langle B\left(
\widehat{v}^{\lambda}+\lambda z,\widehat{v}^{\lambda}\right)  ,q^{\lambda
}\right\rangle _{H}\right|  \\
& \leq C\left\|  \widehat{v}^{\lambda}\right\|  _{V}\left|  q^{\lambda
}\right|  _{H}\left(  \left|  \widehat{v}^{\lambda}\right|  _{H}%
+|\lambda|\left|  z\right|  _{H}\right)  \\
& \leq\frac{\nu}{2}\left\|  \widehat{v}^{\lambda}\right\|  _{V}^{2}%
+C_{2}\left(  \omega\right)  \left|  \widehat{v}^{\lambda}\right|  _{H}%
^{2}+\frac{\lambda^{2}}{\nu}C_{2}\left(  \omega\right)\, ,
\end{align*}
for some constant $C_{2}\left(  \omega\right)  $, and
\begin{align*}
\left|  \left\langle B\left(  \rho^{\lambda},\rho^{\lambda}\right)
,\widehat{v}^{\lambda}\right\rangle _{H}\right|   &  =\left|  \left\langle
B\left(  \widehat{v}^{\lambda}+\lambda z,\lambda z\right)  ,\widehat
{v}^{\lambda}\right\rangle _{H}\right|  \\
&  \leq C|\lambda|\left\|  z\right\|  _{V}\left|
\widehat{v}^{\lambda}\right| _{H}\left(  \left|
\widehat{v}^{\lambda}\right|  _{H}+|\lambda|\left|
z\right|  _{H}\right)  \\
&  \leq|\lambda| C_{3}\left(  \omega\right)  \left|
\widehat{v}^{\lambda }\right|  _{H}\left(  \left|
\widehat{v}^{\lambda}\right|  _{H}+|\lambda|
C_{3}\left(  \omega\right)  \right)  \\
&  \leq|\lambda| C_{3}\left(  \omega\right)  \left|
\widehat{v}^{\lambda }\right|  _{H}^{2}+|\lambda|^{3}C_{3}\left(
\omega\right)\, ,
\end{align*}
for some constant $C_{3}\left(  \omega\right)  $. Hence, from the
equation in weak form for $\widehat{v}^{\lambda}$ (and similarly to
the proof of step 2 above) we deduce, for $\lambda\in\left[
-1,1\right] $,
\begin{align*}
& \frac{1}{2}\left|  \widehat{v}^{\lambda}\left(  t\right)  \right|  _{H}%
^{2}+\frac{\nu}{2}\int_{0}^{t}\left\|  \widehat{v}^{\lambda}\left(  s\right)
\right\|  _{V}^{2}ds\\
& \leq\frac{1}{2}\left|  \lambda w_{0}\right|
_{H}^{2}+\int_{0}^{t}\left( C_{4}\left(  \omega\right)  \left|
\widehat{v}^{\lambda}\left(  s\right) \right|
_{H}^{2}+\lambda^{2}\left(  1+\frac{1}{\nu}\right)  C_{4}\left(
\omega\right)  \right)  ds \, ,
\end{align*}
for some constant $C_{4}\left(  \omega\right)  $.
By Gronwall lemma we get
\[
\lim_{\lambda\longrightarrow0}\sup_{\widetilde{u}_{0}\in B}\sup_{0\leq t\leq
T}\left|  \widehat{v}^{\lambda}(t,\omega,\widetilde{u}_{0})\right|  _{H}^{{}%
}=0 \,,
\]
which implies the claim of this step.

\textbf{Step 4:} (convergence of $w^{\lambda}$). With the notation
$\xi^{\lambda}(t,\omega,\widetilde{u}_{0}):=w^{\lambda}(t,\omega,\widetilde
{u}_{0})-w^{0}(t,\omega,\widetilde{u}_{0})$, let us prove that
\[
\lim_{\lambda\longrightarrow0}\sup_{\widetilde{u}_{0}\in B}\sup_{0\leq t\leq
T}\left|  \xi^{\lambda}(t,\omega,\widetilde{u}_{0})\right|  _{H}^{{}}=0.
\]
We have
\begin{align*}
& \left\langle \xi^{\lambda}(t),\psi\right\rangle _{H}+\int_{0}^{t}%
\nu\left\langle \xi^{\lambda}(s),A\psi\right\rangle _{V,V^{\prime}}ds\\
& =\int_{0}^{t}\left\langle B\left(  u^{0}(s),w^{0}(s)\right)
,\psi
\right\rangle _{H}ds\\
& -\int_{0}^{t}\left\langle B\left(
u^{\lambda}(s),w^{\lambda}(s)\right) +\lambda B\left(
w^{\lambda}(s),w^{\lambda}(s)\right)  ,\psi\right\rangle
_{H}ds\\
& =\int_{0}^{t}\left\langle B\left(  q^{0}(s),w^{0}(s)\right)
,\psi \right\rangle _{H}ds-\int_{0}^{t}\left\langle B\left(
q^{\lambda
}(s),w^{\lambda}(s)\right)  ,\psi\right\rangle _{H}ds\\
& =-\int_{0}^{t}\left\langle B\left(
\rho^{\lambda}(s),w^{0}(s)\right) ,\psi\right\rangle
_{H}ds-\int_{0}^{t}\left\langle B\left(  q^{\lambda
}(s),\xi^{\lambda}(s)\right)  ,\psi\right\rangle _{H}ds \,,
\end{align*}
for all $t\geq0$ and $\psi\in V$. As in the previous steps we deduce
that
\begin{align*}
& \frac{1}{2}\left|  \xi^{\lambda}\left(  t\right)  \right|  _{H}^{2}+\nu
\int_{0}^{t}\left\|  \xi^{\lambda}\left(  s\right)  \right\|  _{V}^{2}ds\\
& \leq\int_{0}^{t}\left|  \left\langle B\left(  \rho^{\lambda}(s),w^{0}%
(s)\right)  ,\xi^{\lambda}\left(  s\right)  \right\rangle _{H}\right|  ds\\
& +\int_{0}^{t}\left|  \left\langle B\left(  q^{\lambda}(s),\xi^{\lambda
}(s)\right)  ,\xi^{\lambda}\left(  s\right)  \right\rangle _{H}\right|  ds\\
& \leq C\int_{0}^{t}\left|  w^{0}\left(  s\right)  \right|
_{H}\left\| \xi^{\lambda}\left(  s\right)  \right\|  _{V}\left|
\rho^{\lambda}(s)\right| _{H}ds.
\end{align*}
Hence,
\[
\frac{1}{2}\left|  \xi^{\lambda}\left(  t\right)  \right|  _{H}^{2}+\frac{\nu
}{2}\int_{0}^{t}\left\|  \xi^{\lambda}\left(  s\right)  \right\|  _{V}%
^{2}ds\leq\frac{C}{\nu}\int_{0}^{t}\left|  w^{0}\left(  s\right)  \right|
_{H}^{2}\left|  \rho^{\lambda}(s)\right|  _{H}^{2}ds.
\]
This implies the claim of this step.

\textbf{Step 5:} (convergence of $u^{\lambda}$). We simply notice that
\[
\left|  u^{\lambda}(t)-u^{0}(t)\right|  _{H}=\left|  q^{\lambda}%
(t)-q^{0}(t)-\lambda w^{\lambda}(t)\right|  _{H}\,,%
\]
therefore, by the results of steps 3 and 4, we have
\[
\lim_{\lambda\longrightarrow0}\sup_{\widetilde{u}_{0}\in B}\sup_{0\leq t\leq
T}\left|  u^{\lambda}(t,\omega,\widetilde{u}_{0})-u^{0}(t,\omega,\widetilde
{u}_{0})\right|  _{H}^{{}}=0.
\]
The proof is complete.
\end{proof}

\section{Random dynamical systems}

In this section we recall few definitions from the theory of random
dynamical systems. For general notions and results see
\cite{Arnold}, and  see \cite{CKS} for analogous concept for
non-autonomous dynamical systems. Here we mainly refer to specific
notions from \cite{crauel-flandoli}.

\subsection{The basic set-up}
Recall from the previous section, the following notation: let
$(\Omega,\mathcal{F},P)$ be a probability space and $\left\{
\theta _{t}:\Omega\longmapsto\Omega,\ t\in\mathbb{R}\right\}  $ a
family of measure preserving transformations such that
$(t,\omega)\longmapsto\theta_{t}(\omega)$ is measurable,
$\theta_{0}=Id$, $\theta_{t+s}=\theta_{t}\circ\theta_{s},$ for
$s,t\in\mathbb{R}$. The flow $(\theta_{t})_{t\in\mathbb{R}}$
together with the probability space $(\Omega,\mathcal{F},P)$ is
called a measurable dynamical system. Furthermore, we suppose that
the map $\theta_{t}$ is ergodic.

\begin{definition}
Let $(X,d)$ be a Polish space (i.e. complete separable metric
space) and $\mathcal{B}$ its Borel $\sigma$-algebra. Let
$\mathbb{R}^{+}=[0,\infty)$. A map

\begin{align*}
\varphi:\mathbb{R}^{+}\times\Omega\times X  &  \longmapsto X\\
(t,\omega,x)  &  \longmapsto\varphi(t,\omega)x
\end{align*}
is called a measurable {\bf random dynamical system} (RDS) on $X$
over $(\Omega,\mathcal{F},P,\theta_{t})$ if the following
properties are satisfied
\begin{enumerate}

\item $\varphi$ is
$(\mathcal{B}(\mathbb{R}^{+})\bigotimes\mathcal{F}\bigotimes
\mathcal{B}, \mathcal{B})$ measurable, where
$\mathcal{B}(\mathbb{R}^{+})$ is the Borel $\sigma$-algebra on
$\mathbb{R}^{+}$;

\item
\begin{equation}\label{cocycle_property}
\varphi(t+s,\omega)=\varphi(t,\theta_{s}\omega)\circ\varphi(s,\omega),
\end{equation}
for all $t,\ s\in \mathbb{R}^{+}$ and $\varphi(0,\omega)=Id$, for
all $\omega\in \Omega$. Property (\ref{cocycle_property}) is called
the {\bf Cocycle property}.
\end{enumerate}
\end{definition}

An RDS $\varphi$ is said to be continuous or differentiable if for
every fixed $(t,\omega)\in \mathbb{R}^{+}\times \Omega$,
$\varphi(t,\omega) :X\longmapsto X$ is continuous or
differentiable respectively.

Instead of assuming
(\ref{cocycle_property}) for all $\omega\in \Omega$, it suffices
to assume it for all $\omega$ from a measurable
$\theta_{t}$-invariant subset of full measure.

\subsection{Attraction, absorption and invariance}
Let $(X,d)$ be a metric space. For two nonempty sets $A, B\subset
X$ , we recall the Hausdorf semi-metric $d_{H}\left( A,B\right)
=\sup_{x\in A}\inf_{y\in B}d\left( x,y\right)$.

We observe that $d_{H}$ restricted to the family of all nonempty
closed subsets of $X$ is a metric, see \cite{Castaing-Valadier}.

\begin{definition}\label{definition_random_set}
Let $(\Omega,\mathcal{F})$ be a measurable space and let $(X,d)$ be
a Polish space. A set valued map $K:\Omega\longrightarrow 2^{X}$,
taking values in the closed subsets of $X$, is said to be measurable
if for each fixed $x\in X$, the map $\omega\longmapsto
d_{H}(x,K(\omega))$ is measurable. The map $K$ is often called a
{\bf closed random set}.
\end{definition}

\begin{definition}\label{definition_invariant_set} Let $\varphi:
\mathbb{R}^{+}\times\Omega\times X$ such that
$(t,\omega,x)\longmapsto\varphi(t,\omega)x\in X$ be a measurable RDS
on a Polish space $(X,d)$ over a measurable dynamical system
$(\Omega,\mathcal{F},P,\theta_{t})$. A closed random set $K$ is
called $\varphi$-{\bf forward invariant} if for all $\omega\in
\Omega$,
\begin{equation}\label{invariant}
\varphi(t,\omega)K(\omega)\subseteq K(\theta_{t}\omega)\, \, \text{
for all} \,\, \, t>0\,.
\end{equation}

A closed random set $K$ is called {\bf strictly} $\varphi$-{\bf
forward invariant} if  for all $\omega\in \Omega$,
\begin{equation}\label{strictly_invariant}
\varphi(t,\omega)K(\omega)=K(\theta_{t}\omega)\, \, \text{ for all}
\,\, \,  t>0.
\end{equation}
\end{definition}

\begin{remark} By substituting $\theta_{-t}\omega$ for $\omega$
in Definition \ref{definition_invariant_set}, we get the following
equivalent version of Definition \ref{definition_invariant_set}. A
closed random set $K$ is called {\bf strictly} $\varphi$-{\bf
forward invariant} if for all for all $\omega\in \Omega$,
\begin{equation}\label{invariant2}
\varphi(t,\theta_{-t}\omega)K(\theta_{-t}\omega)\subseteq
K(\omega)\, \, \text{ for  all}\, \,\,  t>0
\end{equation}
or respectively

\begin{equation}\label{strictly_invariant2}
\varphi(t,\theta_{-t}\omega)K(\theta_{-t}\omega)= K(\omega) \, \,
\text{ for all} \,\, \, t>0.
\end{equation}
\end{remark}

\begin{definition}\label{definition_absorbing_set}
A closed random set $K$ is said to absorb the set $B\subset X$, $B$
is fixed non-random,  if there exists a random variable
$t_{B}(\omega) $ such that, for $P$-a.e. $\omega\in\Omega$,
\begin{equation}\label{absorbtion_property}
\varphi(t,\theta_{-t}\omega)B\subset K(\omega)\quad\text{for all }%
t>t_{B}(\omega).
\end{equation}
The smallest $t_{B}(\omega)\geq 0$ for which
(\ref{absorbtion_property}) holds is called the random absorption
time of $B$ by $K$.
\end{definition}

\begin{remark}
Note that $\varphi(t,\theta_{-t}\omega)x$ can be thought of as the
position of the trajectory at time 0, which was in $x$ at time $-t$.
\end{remark}

\begin{definition}\label{definition_omega_limit_set}
For a given closed random set $K$, the $\omega${\bf-limit set} of
$K$ is defined to be the random set

\begin{equation}\label{omega_limit_set}
\Lambda_{K}(\omega)=\bigcap_{n\geq0}\overline{\bigcup_{t\geq
n}\varphi (t,\theta_{-t}\omega)K(\theta_{-t}\omega)}.
\end{equation}
\end{definition}

\begin{remark}
\begin{enumerate}
\item {\it A priori} $\Lambda_{K}(\omega)$ can be an empty set.

\item We have the following equivalent version of Definition
\ref{definition_omega_limit_set}:

$$\Lambda_{K}(\omega)=\left\{ y\in X:\ \exists t_{n}\rightarrow\infty,\
\left\{x_{n}\right\}\subset K(\theta_{-t_{n}}\omega),\
\lim_{n\rightarrow\infty}\varphi(t_{n},\theta_{-t_{n}}\omega)x_{n}=y\right\}.$$

\item Since $\overline{\bigcup_{t\geq n}\varphi
(t,\theta_{-t}\omega)K(\theta_{-t}\omega)}$ is closed, then
$\Lambda_{K}(\omega)$ is closed as well.
\end{enumerate}
\end{remark}

\begin{definition}\label{definition_attractor}
A random set $\mathcal{A}(\omega)$ is called a random attractor
associated with the random dynamical system $\varphi$ if, for
$P$-a.e. $\omega\in\Omega$, the following is satisfied:
\begin{enumerate}

\item $\mathcal{A}(\omega)$ is a nonempty compact subset of $X$,

\item
$\varphi(t,\omega)\mathcal{A}(\omega)=\mathcal{A}(\theta_{t}\omega
),\ \forall\ t\geq0$,

\item for every $B\subset X$ bounded (and non-random)
\[
\lim_{t\longrightarrow\infty}d_{H}\left(
\varphi(t,\theta_{-t}\omega )B,\mathcal{A}(\omega)\right)  =0
\]
\end{enumerate}
\end{definition}

The following theorem about the existence of random attractors is
due to Crauel and Flandoli \cite{crauel-flandoli}.

\begin{theorem}\label{thm_flandoli_crauel}
Suppose there exists a closed random set $D$ which is absorbing
every bounded non-random set $B\subset X$, and for which $D(\omega)$
is a compact subset of $X$ for $P$-a.e. $\omega\in\Omega$. Then, the
set
\[
\mathcal{A}(\omega)=\overline{\bigcup_{B\subset
X}\Lambda_{B}(\omega)}\,,%
\]
is a random attractor for $\varphi$. Where the union above is taken
over all the bounded  and non-random $B\subset X$,  and
$\Lambda_{B}(\omega)$ is the $\omega$-limit set of $B$.
\end{theorem}

\begin{remark}
In Crauel \cite{crauel-unique-attractors} it is shown that, under
the ergodicity assumption on $\theta_{t}$, there exists a compact
set $K(\omega)\subset X$ such that, for $P$-a.e.
$\omega\in\Omega$, the random attractor
is the $\omega -$limit set of $K(\omega)$, that is,%

\[
\mathcal{A}(\omega)=\bigcap_{n\geq0}\overline{\bigcup_{t\geq
n}\varphi (t,\theta_{-t}\omega)K(\omega)}.
\]
\end{remark}

\subsection{Random attractor dimensionality}

We are interested in the property of finite Hausdorff dimensionality
of the random attractor. The two most relevant results for this
purpose are the works of Debussche \cite{debussche1},
\cite{debussche2}. We apply the result of the second one of these
papers, based on a property called \textit{random squeezing
property}, which was inspired by the \textit{squeezing property} in
the deterministic case that was introduced in \cite{Foias-Temam}
(see, also, \cite{Constantin-Foias-Nicolaenko-Temam} and
\cite{Ladyzhenskaya}).  This property is also used in  the proof of
finite number of determining modes. The fact that the random
attractor is not uniformly bounded makes the corresponding random
squeezing property depend exponentially on a random variable.
However, an ergodic argument will make it possible to work with this
weaker property.

\begin{definition}\label{definition_debussche}
(\cite{debussche1}) Let $H$ be a separable Hilbert space with norm
$\left| .\right|  _{H}$, $\varphi(t,\omega)$ a random dynamical
system in $H$ with random attractor $\mathcal{A}(\omega)$. We say
that $\varphi(t,\omega)$ satisfies the random squeezing property if
there exist a random variable $C_{5}(\omega)$, a finite-dimensional
projector $\Pi$ in $H$, and positive
numbers $\mu,\delta$ such that, for $P$-a.e. $\omega\in\Omega$,%

\begin{equation}
|\Pi\varphi(t,\omega)u_{0}-\Pi\varphi(t,\omega)v_{0}|_{H}\leq
e^{\int_{0}^{t}C_{5}(\theta_{s}\omega)ds}|u_{0}-v_{0}|_{H}%
\end{equation}
and%

\begin{equation}
|(I-\Pi)\left(  \varphi(t,\omega)u_{0}-\varphi(t,\omega)v_{0}\right)
|_{H}\leq\left(  e^{-\mu
t}+\delta e^{\int_{0}^{t}C_{5}(\theta_{s}\omega)ds}\right)
|u_{0}-v_{0}|_{H}%
\end{equation}
for every $t\geq0$, and every $u_{0},v_{0}\in\mathcal{A}(\omega)$.
\end{definition}

\begin{theorem}\label{theorem_debussche}
(\cite{debussche1}) There exist absolute constants
$K_{1},K_{2},K_{3}$ such that if $\varphi(t,\omega)$ is  a random
dynamical system that satisfies:

(i) the random squeezing property, mentioned in Definition
\ref{definition_debussche}, with a random variable $C_{5}(\omega)$,
a finite-dimensional projector $\Pi$ and two positive numbers
$\mu,\delta$,

and

(ii)  the expected value with respect to the measure $P$
\[
E(C_{5}(\omega))<\infty,\ \delta\leq K_{1}\ \text{and}\ \mu\geq K_{2}%
E\left(C_{5}(\omega)\right),
\]
then, for $P$-a.s.
$\omega\in\Omega$, the random attractor $\mathcal{A}%
(\omega)$ of $\varphi(t,\omega)$ has finite Hausdorff dimension
which is less than $K_{3}R(\Pi )\log{R(\Pi)}$, where $R(\Pi)$ is the
rank of the projector $\Pi$.
\end{theorem}

\section{Application to the Shell model}

In Theorem \ref{theorem_goy_existence} we have constructed, for
every $\lambda\in \mathbb{R}$, a random dynamical system
$\varphi^{\lambda}(t,\omega)$ associated with equation
(\ref{GOY_lambda}). In this section we prove the existence of the
random attractor associated with $\varphi^{\lambda}(t,\omega)$. At
the end of the section we also prove that the random attractor, as a
function of $\lambda$, is upper semi-continuous. For the upper
semi-continuity of deterministic attractors with respect to a
parameter see, e.g., \cite{Hale}.

\subsection{Auxiliary problem}
As in step 2 of section 2.2, we will introduce an auxiliary
Ornstein-Uhlenbeck process.  This is a slightly different process
but will have mainly the same properties as the process introduced
in step 2 of section 2.2. Following the steps of section 2, let
$\Omega_{W}^{0}\in \mathcal{F}$ introduced in section 2. Let
$\alpha>0$ be an arbitrary constant. For $\omega\in\Omega_{W}^{0}$,
let
\begin{equation}
\widetilde{z}^{\alpha}(t,\omega)=\int_{-\infty}^{t}(\nu\widetilde{A}%
+\alpha I)e^{-(\nu\widetilde{A}%
+\alpha
I)(t-s)}(\widetilde{W}(t,\omega)-\widetilde{W}(s,\omega))ds
\end{equation}
which is well defined and bounded in $\widetilde{V}$.

The process $\widetilde{z}^{\alpha}(t)$, $t\in \mathbb{R}$ is a
Gaussian, stationary and ergodic process. Moreover, it is a solution
of the equation
\begin{equation}\label{z-equation}
d\widetilde{z}^{\alpha}(t)=(-\nu
\widetilde{A}-\alpha)\widetilde{z}^{\alpha}(t)dt+d\widetilde{W}(t),
\end{equation}
i.e. for all $t\in \mathbb{R}$ and $P-$a.s.

\begin{equation}
\widetilde{z}^{\alpha}(t)=\int_{-\infty}^{t}e^{-(\nu\widetilde{A}%
+\alpha)(t-s)}d\widetilde{W}(s).
\end{equation}
In particular, for each component $\widetilde{z}^{\alpha}_{n}$ one
has
\begin{equation}
\widetilde{z}^{\alpha}_{n}(t)=\int_{-\infty}^{t}e^{-(\nu k_{n}^{2}%
+\alpha)(t-s)}\sigma_{n}d\widetilde{\beta}_{n}(s).
\end{equation}
Moreover,

\begin{align}\label{z_alpha_estimate1}
E\|\widetilde{z}^{\alpha}(t)\|_{\tilde{V}}^{2} =
E|\widetilde{A}^{1/2}\widetilde{z}^{\alpha}(t)|_{\tilde{H}}^{2}&
=E\sum_{n=1}^{\infty}(k_{n}\widetilde
{z}^{\alpha}_{n}(t))^{2}\nonumber\\
&  =E\sum_{n=1}^{\infty}\left(  \int_{-\infty}^{t}k_{n}e^{-(\nu k_{n}^{2}%
+\alpha)(t-s)}\sigma_{n}d\widetilde{\beta}_{n}(s)\right)  ^{2}\nonumber\\
&  \leq\sum_{n=1}^{\infty}E\left(  \int_{-\infty}^{t}k_{n}e^{-(\nu k_{n}^{2}%
+\alpha)(t-s)}\sigma_{n}d\widetilde{\beta}_{n}(s)\right)  ^{2}\nonumber\\
&  =\sum_{n=1}^{\infty}\int_{-\infty}^{t}k_{n}^{2}e^{-2(\nu
k_{n}^{2}+\alpha
)(t-s)}\sigma^{2}_{n}ds\nonumber\\
&  =\sum_{n=1}^{\infty}\frac{k_{n}^{2}\sigma^{2}_{n}}{2(\nu
k_{n}^{2}+\alpha)}<\infty\,.%
\end{align}

Furthermore, $E\|\widetilde{z}^{\alpha}(t)\|_{\tilde{V}}^{2}$ tends
to 0, when $\alpha\rightarrow\infty$. In particular, there exists
$\alpha_* (\nu)>0$ such that
\begin{equation}\label{E-z}
E\|\widetilde{z}^{\alpha}(t)\|_{\tilde{V}}^{2} \le
\frac{k_0\nu}{8C_*} \quad \text{for all} \quad \alpha\ge \alpha_*\,,
\end{equation}
where $C_*$ is the constant in the inequality (\ref{V-prime}).
\subsection{Absorbing compact set}\label{absorbing-set}
Let $\omega\in \Omega_{W}^{0}$ be given and let us introduce the
random differential equation

\begin{align}\label{v_alpha}
\frac{d\widetilde{v}(t,\omega)}{dt}&+\nu\widetilde{A}\widetilde{v}(t,\omega)=\\
\nonumber
&-\widetilde{B}_{\lambda}(\widetilde{v}(t,\omega)+\widetilde{z}^{\alpha}(t,\omega),\widetilde{v}(t,\omega)+\widetilde
{z}^{\alpha}(t,\omega))+ \alpha\widetilde{z}^{\alpha}(t,\omega).
\end{align}
%
Notice that in fact $\widetilde{v}$ depends on $\alpha$, because
$\widetilde{z}^{\alpha}$ depends on $\alpha$.  It is not difficult
to prove, using a Galerkin method, that for each $\omega\in
\Omega_{W}^{0}$ and $t_{0}\in \mathbb{R}$, with
$\widetilde{v}_{t_{0}}(\omega)\in\widetilde{H}$ is given, there
exists a unique solution $\widetilde{v}(t,\omega)$ defined on
$[t_{0},\infty)$ to (\ref{v_alpha}) such that
\begin{equation}\label{v_alpha_initial}
\widetilde{v}(t_{0},\omega)=\widetilde{v}_{t_{0}}(\omega)\,,
\end{equation}
and such that

$$\widetilde{v}(\cdot,\omega)\in
C([t_{0},\infty),\widetilde{H})\bigcap
L^{2}((t_{0},\infty),\widetilde{V}).$$ We refer to
\cite{Fladissipative} for more detailed computations.

Let us define

\[
\varphi^{\lambda}(t,\omega)\widetilde{u}_{t_{0}}=\widetilde{v}\left(
t,\omega\right)  +\widetilde{z}^{\alpha}\left( t,\omega\right)
\]
where $\widetilde{v}$ is the solution of (\ref{v_alpha}) with

$$\widetilde{v}_{t_{0}}(\omega)=\widetilde{u}_{t_{0}}(\omega)-\widetilde{z}^{\alpha}(t_{0},\omega).$$

We would like to prove the existence of a compact absorbing set in
$\widetilde{H}$ at time $t=0$. Through this section we will take $B$
to be a bounded set in $\widetilde{H}$, and that  for any $t_{0}\in
\mathbb{R}$ we will assume  $\widetilde{u}(t_{0})\in B$; moreover,
$\widetilde{v}$ is the  solution of \eqref{v_alpha} and
\eqref{v_alpha_initial} with
$$\widetilde{v}_{t_{0}}(\omega)=\widetilde{u}(t_{0},\omega)-\widetilde{z}^{\alpha}(t_{0},\omega).$$
Let $t_{0}<-1$ and $t\in[-1,0]$.

\begin{lemma}\label{radius_estimate1}
Let $\widetilde{v}_{t_{0}}\in \widetilde{H}$ and $\widetilde{v}$ be
a solution of \eqref{v_alpha} associated with the initial condition
$\widetilde{v}_{t_{0}}$. Then, for all $t_{0}<-1$ and for all
$t\in\lbrack-1,0]$

\begin{equation}
|\widetilde{v}(t)|_{\tilde{H}}^{2}\leq|\widetilde{v}(t_{0})|_{\tilde{H}}^{2}e^{\int_{t_{0}}^{t}\left(
2C_{*}\Vert \widetilde{z}^{\alpha}(s)\Vert_{\tilde{V}}-\frac{k_{0}\nu}{2}\right)  ds}+\int_{t_{0}}%
^{t}f(s)\displaystyle e^{\int_{s}^{t}\left(  2C_{*}\Vert
\widetilde{z}^{\alpha}(r)\Vert_{\tilde{V}}-\frac
{k_{0}\nu}{2}\right)  dr}ds.
\end{equation}
where
\[
f(t):=\frac{4C_{*}^{2}}{\nu}|\widetilde{z}^{\alpha}(t)|_{\tilde{H}}^{4}+\frac{8\alpha^{2}}{k_{0}%
\nu}|\widetilde{z}^{\alpha}(t)|_{\tilde{H}}^{2},
\]
and $C_*$ is the constant in the inequality (\ref{V-prime}).
\end{lemma}

\begin{proof} As before, the proof is formal and can be made rigorous
by applying the Galerkin approximation procedure.
Let us take the inner produce in $\widetilde{H}$ of
equation (\ref{v_alpha}) with  $\widetilde{v}$ to obtain%

\[
<\frac{d\widetilde{v}}{dt},\widetilde{v}>+\nu<\widetilde{A}\widetilde
{v},\widetilde{v}>=-<\widetilde{B}_{\lambda}(\widetilde{v}+\widetilde{z}^{\alpha},
\widetilde{v}+\widetilde{z}^{\alpha}),\widetilde{v}%
>+\alpha<\widetilde{z}^{\alpha},\widetilde{v}>.
\]%
Using Lemma \ref{lemma su B tilde}, inequality (\ref{V-prime}) and
Young's inequality, we estimate the right-hand side of the above and
get

\begin{align}
|<\widetilde{B}_{\lambda}(\widetilde{v}+\widetilde{z}^{\alpha},\widetilde{v}%
+\widetilde{z}^{\alpha},\widetilde{v}>| &
=|<\widetilde{B}_{\lambda}(\widetilde
{v}+\widetilde{z}^{\alpha},\widetilde{v}),\widetilde{v}>+<\widetilde{B}_{\lambda
}(\widetilde{v}+\widetilde{z}^{\alpha},\widetilde{z}^{\alpha},\widetilde{v}>|\nonumber\\
&
=|\widetilde{B}_{\lambda}(\widetilde{v}+\widetilde{z}^{\alpha},\widetilde{z}^{\alpha}
,\widetilde{v}>|=|-\widetilde{B}_{\lambda}(\widetilde{v}+\widetilde{z}^{\alpha}
,\widetilde{v}),\widetilde{z}^{\alpha}>|\nonumber\\
&
=|-\widetilde{B}_{\lambda}(\widetilde{v},\widetilde{v}),\widetilde{z}^{\alpha}>
-\widetilde{B}_{\lambda}(\widetilde{z}^{\alpha},\widetilde{v}),\widetilde{z}^{\alpha}>|\nonumber\\
& \leq\Vert\widetilde{B}_{\lambda}(\widetilde{v},\widetilde{v})\Vert
_{\widetilde{V}^{\prime}}\Vert\widetilde{z}^{\alpha}\Vert_{\tilde{V}}+|\widetilde{B}_{\lambda
}(\widetilde{z}^{\alpha},\widetilde{v})|_{\widetilde{H}}|\widetilde{z}^{\alpha}|_{\tilde{H}}\nonumber\\
&  \leq C_{*}\left(
|\widetilde{v}|_{\tilde{H}}^{2}\Vert\widetilde{z}^{\alpha}\Vert_{\tilde{V}}+\Vert
\widetilde{v}\Vert_{\tilde{V}}|\widetilde{z}^{\alpha}|_{\tilde{H}}^{2}\right)  \nonumber\\
&  \leq C_{*}|\widetilde{v}|_{\tilde{H}}^{2}\Vert\widetilde{z}^{\alpha}\Vert_{\tilde{V}}+\frac{\nu}{2}%
\Vert\widetilde{v}\Vert_{\tilde{V}}^{2}+\frac{2C_{*}^{2}}{\nu}|\widetilde{z}^{\alpha}|_{\tilde{H}}^{4}.
\end{align}
For the other term we apply the Cauchy-Schwarz and Young's
inequalities we have

\begin{equation}
|\alpha<\widetilde{z}^{\alpha},\widetilde{v}>|\leq\alpha|\widetilde{z}^{\alpha}|_{\tilde{H}}|\widetilde
{v}|_{\tilde{H}}\leq\frac{k_{0}\nu}{4}|\widetilde{v}|_{\tilde{H}}^{2}+\frac{4\alpha^{2}}{k_{0}\nu
}|\widetilde{z}^{\alpha}|_{\tilde{H}}^{2}.
\end{equation}
Hence, we get that
\begin{equation}\label{v-intermediate}
\frac{1}{2}\frac{d}{dt}|\widetilde{v}|_{\tilde{H}}^{2}+\frac{\nu}{2}\Vert\widetilde
{v}\Vert_{\tilde{V}}^{2}\leq
C_{*}|\widetilde{v}|_{\tilde{H}}^{2}\Vert
\widetilde{z}^{\alpha}\Vert_{\tilde{V}}+\frac{2C_{*}^{2}}{\nu
}|\widetilde{z}^{\alpha}|_{\tilde{H}}^{4}+\frac{k_{0}\nu}{4}|\widetilde{v}|_{\tilde{H}}^{2}+\frac{4\alpha^{2}%
}{k_{0}\nu}|\widetilde{z}^{\alpha}|_{\tilde{H}}^{2}.
\end{equation}
Using the Poincar\'{e}-like inequality
$\nu\Vert\widetilde{v}\Vert_{\tilde{V}}^{2}\geq\nu k_{0}|\widetilde
{v}|_{\tilde{H}}^{2}$, we get%

\[
\frac{d}{dt}|\widetilde{v}|_{\tilde{H}}^{2}\leq
2C_{*}|\widetilde{v}|_{\tilde{H}}^{2}\Vert\widetilde{z}^{\alpha}
\Vert_{\tilde{V}}+\frac{4C_{*}^{2}}{\nu}|\widetilde{z}^{\alpha}|_{\tilde{H}}^{4}-\frac{k_{0}\nu}%
{2}|\widetilde{v}|_{\tilde{H}}^{2}+\frac{8\alpha^{2}}{k_{0}\nu}|\widetilde{z}^{\alpha}|_{\tilde{H}}^{2}.
\]
We integrate over $(t_{0},t)$ and get%

\[
|\widetilde{v}(t)|_{\tilde{H}}^{2}\leq|\widetilde{v}(t_{0})|_{\tilde{H}}^{2}+\int_{t_{0}}^{t}\left(
2C_{*}\Vert\widetilde{z}^{\alpha}(r)\Vert_{\tilde{V}}-\frac{k_{0}\nu}{2}\right)
|\widetilde {v}(r)|_{\tilde{H}}^{2}dr+\int_{t_{0}}^{t}f(r)dr.
\]

Using Gronwall Lemma we get the result.
\end{proof}

\begin{lemma}\label{radius_estimate2}
Let $\alpha \ge \alpha_*(\nu)$,  such that~(\ref{E-z}) holds.
Suppose that the assumptions of Lemma \ref{radius_estimate1} are
satisfied, and let $f$ be given by Lemma \ref{radius_estimate1}. Let
$$\displaystyle
R_{1}^{t}(\omega):= 1+\int_{-\infty}^{t}f(s)e^{\int_{s}^{t}\left(
2C_{*}\Vert\widetilde{z}^{\alpha}(r)\Vert_{\tilde{V}}-\frac{k_{0}\nu}{2}\right)dr}ds$$
then, the ball $B(0,R_{1}^{0}(\omega))\subset\widetilde{H}$ is an
absorbing set at time $t\in [-1,0]$, for the system (\ref{v_alpha}).
\end{lemma}

Notice that in principle $R_{1}^{t}(\omega)$ depends on $\alpha$,
however, if we fix $\alpha = \alpha_{0}:= 2 \alpha_*(\nu)$  (see
(\ref{E-z}))  then $R_{1}^{t}(\omega)$  will depend only on $\nu$
and the statement will still be valid.

\begin{proof}
Using the ergodicity of the process $\widetilde{z}^{\alpha}$ we have that%

\[
-\frac{1}{t_{0}}\int_{t_{0}}^{0}\Vert\widetilde{z}^{\alpha}(s)\Vert_{\tilde{V}}
ds\longrightarrow
_{t_{0}\rightarrow-\infty}E\Vert\widetilde{z}^{\alpha}(0)\Vert_{\tilde{V}},\,\,
P-\text{a.s.}\,,
\]
Hence, there exists $s_{0}(\omega) <0$ such that for every $t_{0}\leq s_{0}%
(\omega)$,
\begin{equation}\label{s0}
-\frac{1}{t_{0}}\int_{t_{0}}^{0}\Vert\widetilde{z}^{\alpha}(s)\Vert_{\tilde{V}}
ds\leq 2E\Vert\widetilde{z}^{\alpha}(0)\Vert_{\tilde{V}}.
\end{equation}
Then,

\[
\exp{\left( t_{0}\left(  \frac{k_{0}\nu}{2}+\frac{1}{t_{0}}\int_{t_{0}%
}^{0}2C_{*}\Vert\widetilde{z}^{\alpha}(s)\Vert_{\tilde{V}} ds\right)
\right) }\leq \exp{ \left( t_{0}\left(
\frac{k_{0}\nu}{2}-2C_{*}E\Vert\widetilde{z}^{\alpha}(0)\Vert_{\tilde{V}}\right)\right)
}.
\]
Moreover, thanks to \eqref{z_alpha_estimate1} and (\ref{E-z})  we
have for all $\alpha \ge \alpha_*$
\begin{equation}\label{E-alpha}
E\Vert\widetilde{z}^{\alpha}(0)\Vert_{\tilde{V}}<\frac{k_{0}\nu}{8C_{*}}.
\end{equation}
Hence, for every $t_{0}\leq s_{0}(\omega)$ and $\alpha\ge \alpha_*$
one has
\[
\exp{\left(t_{0}\left(  \frac{k_{0}\nu}{2}+\frac{1}{t_{0}}\int_{t_{0}%
}^{0}2C_{*}\Vert\widetilde{z}^{\alpha}(s)\Vert_{\tilde{V}}
ds\right)\right) }\leq \exp{\left(
t_{0}\frac{k_{0}\nu}{8C_{*}}\right)}\,.
\]

In addition, see \cite{Fladissipative} for more details, one can easily prove that there exists an a.s. finite
random constant $C_{7}(\omega)$  such that

\begin{equation}\label{sub_exponential_growth}
|\widetilde{z}^{\alpha}(t)|_{\tilde{H}}\leq C_{7}(\omega)|t|,\, \, {\rm for}\ {\rm all}\, \, t\leq -1.
\end{equation}

Let us assume that the bounded ball $B\subset \tilde{H}$ is inside a ball of radius $\rho_{1}$, then
$$|\widetilde{u}(t_{0})|_{\tilde{H}}\leq \rho_{1},\ {\rm for}\ {\rm all}\ t_{0}\leq 0.$$
Hence, for every $t_{0}\leq s_{0}(\omega)$ and $\alpha\ge \alpha_*$

\begin{eqnarray*}
|\widetilde{v}(t)|_{\tilde{H}}^{2}&\leq&|\widetilde{u}(t_{0})|_{\tilde{H}}^{2}\exp{\left(
t_{0}\frac{k_{0}\nu}{8C_{*}}\right)}+|\widetilde{z}^{\alpha}(t_{0}|_{\tilde{H}}^{2}\exp{\left(
t_{0}\frac{k_{0}\nu}{8C_{*}}\right)}\\
&+&\int_{t_{0}} ^{t}f(s)\displaystyle e^{\int_{s}^{t}\left(
2C_{*}\Vert \widetilde{z}^{\alpha}(r)\Vert_{\tilde{V}}-\frac
{k_{0}\nu}{2}\right)  dr}ds\\
&\leq& \left(\rho_{1}^{2}+C_{7}^{2}(\omega)|t_{0}|^{2}\right)\exp{\left(
t_{0}\frac{k_{0}\nu}{8C_{*}}\right)}\\
&+&\int_{-\infty} ^{t}f(s)\displaystyle e^{\int_{s}^{t}\left(
2C_{*}\Vert \widetilde{z}^{\alpha}(r)\Vert_{\tilde{V}}-\frac
{k_{0}\nu}{2}\right)  dr}ds
\end{eqnarray*}
Now choose $s_{1}(\omega)<0$ such that
$\left(\rho_{1}^{2}+C_{7}^{2}(\omega)|t_{0}|^{2}\right)\exp{\left(
t_{0}\frac{k_{0}\nu}{8C_{*}}\right)}\leq 1$, for all $t_0 \le
s_{1}(\omega)$, and let us denote by
$t_{B}(\omega)=\min\left\{s_{0}(\omega),s_{1}(\omega)\right\}$, then
we get that the integral inside $R_{1}^{t}(\omega)$ is a.s.
convergent and that for every $t\leq t_{B}(\omega)$
$$|\widetilde{v}(t)|_{\tilde{H}}^{2}\leq R_{1}^{t}(\omega).$$ Hence, the ball $B(0,R_{1}^{t}(\omega))$ is an absorbing ball at time $t$. This completes the proof.
\end{proof}

\begin{lemma}\label{radius_estimate3}
Suppose  that the assumptions of Lemma \ref{radius_estimate1} are
satisfied, and assume that $\alpha\ge \alpha_*$ and $|s_0(\omega)|$
is large enough such that~(\ref{E-alpha}) and~(\ref{s0}) hold,
respectively. In addition, assume that $\widetilde{v}_{t_{0}}\in
\widetilde{V}$, then for every $t_{0}\leq t_{B}(\omega)$, there
exists $R_{2}(\omega)$, $P$- a.s. bounded, such that
\[
\int_{-1}^{0}\Vert\widetilde{v}(t)\Vert_{\tilde{V}}^{2}dt\leq
R_{2}(\omega),
\]
where
\[
R_{2}(\omega):=\frac{1}{\nu}R_{1}^{\{t=-1\}}(\omega)+\int_{-1}^{0}f(t)dt+
\int_{-1}^{0}R_{1}^{s}(\omega)\left(2C_{*}\|\widetilde{z}^{\alpha}(s)\|_{\tilde{V}}+\frac{k_{0}\nu}{2}\right)ds.
\]
\end{lemma}

Notice again that in principle $R_{2}(\omega)$ depends on
$\alpha$, however, if we fix $\alpha = \alpha_{0}:= 2 \alpha_*(\nu)$  then
$R_{2}(\omega)$  will depend only on $\nu$ and the statement of Lemma \ref{radius_estimate3}
will still be valid.

\begin{proof}
Integrate (\ref{v-intermediate}) over $(-1,0)$, then use Lemma
\ref{radius_estimate1} to estimate
$|\widetilde{v}(-1)|_{\tilde{H}}^{2}$.
\end{proof}

\begin{lemma} \label{radius_estimate4} Assume the assumptions of Lemma \ref{radius_estimate3}
are satisfied, then there exists $R_{3}(\omega)$, $P$-a.s. finite such
that
\[
\Vert\widetilde{v}(0)\Vert_{\tilde{V}}^{2}\leq R_{3}(\omega),
\]
where
\begin{eqnarray*}
R_{3}(\omega)&:=&\left(  R_{2}(\omega)+\int_{-1}^{0}\left(  2C_{*}^{2}(R_{1}%
^{s}(\omega)+|\widetilde{z}^{\alpha}(s)|_{\tilde{H}}^{2})\Vert\widetilde{z}^{\alpha}(s)\Vert_{\tilde{V}}^{2}%
+\frac{2\alpha^{2}}{\nu}|\widetilde{z}^{\alpha}(s)|_{\tilde{H}}^{2}\right)  ds\right)\\
& &\times \exp({\int
_{-1}^{0}2C_{*}^{2}(R_{1}^{s}(\omega)+|\widetilde{z}^{\alpha}(s)|_{\tilde{H}}^{2})ds)}\,.
\end{eqnarray*}
\end{lemma}

\begin{proof}
Let us take the inner product in $\widetilde{H}$ of the
equation~(\ref{v_alpha}) with $\widetilde{A}\widetilde{v}$, we get

\begin{align*}
\frac{1}{2}\frac{d}{dt}\Vert\widetilde{v}\Vert_{\tilde{V}}^{2}+\nu|\widetilde{A}%
\widetilde{v}|_{\tilde{H}}^{2} &
=-<\widetilde{B}_{\lambda}(\widetilde{v}+\widetilde
{z}^{\alpha},\widetilde{v}+\widetilde{z}^{\alpha}),\widetilde{A}\widetilde{v}>+\alpha
<\widetilde{z}^{\alpha},\widetilde{A}\widetilde{v}>\\
&  \leq C_{*}|\widetilde{B}_{\lambda}(\widetilde{v}+\widetilde{z}^{\alpha}%
,\widetilde{v}+\widetilde{z}^{\alpha})|_{\tilde{H}}|\widetilde{A}\widetilde{v}|_{\tilde{H}}+\alpha
|\widetilde{z}^{\alpha}|_{\tilde{H}}|\widetilde{A}\widetilde{v}|_{\tilde{H}}\\
&
\leq\frac{\nu}{2}|\widetilde{A}\widetilde{v}|_{\tilde{H}}^{2}+C_{*}^{2}|\widetilde
{v}+\widetilde{z}^{\alpha}|_{\tilde{H}}^{2}\Vert\widetilde{v}+\widetilde{z}^{\alpha}\Vert_{\tilde{V}}^{2}+\frac
{4\alpha^{2}}{\nu}|\widetilde{z}^{\alpha}|_{\tilde{H}}^{2}\, .%
\end{align*}
In the above estimate, we have used Lemma \ref{lemma su B tilde} and the Young's inequality.
Hence,
\begin{eqnarray*}
\Vert\widetilde{v}(t)\Vert_{\tilde{V}}^{2}&\leq&\Vert\widetilde{v}(s)\Vert_{\tilde{V}}^{2}+\int_{s}%
^{t}2C_{*}^{2}|\widetilde{v}(r)+\widetilde{z}^{\alpha}(r)|_{\tilde{H}}^{2}\Vert\widetilde{v}%
(r)\Vert_{\tilde{V}}^{2}dr\\
&+&\int_{s}^{t}\left( 2C_{*}^{2}|\widetilde{v}(r)+\widetilde
{z}^{\alpha}(r)|_{\tilde{H}}^{2}\Vert\widetilde{z}^{\alpha}(r)\Vert_{\tilde{V}}^{2}+\frac{2\alpha^{2}}{\nu}%
|\widetilde{z}^{\alpha}(r)|_{\tilde{H}}^{2}\right)  dr \,.
\end{eqnarray*}
Using Gronwall lemma, we get%

\begin{align*}
\Vert\widetilde{v}(t)\Vert_{\tilde{V}}^{2} &  \leq\Vert\widetilde{v}(s)\Vert_{\tilde{V}}^{2}%
e^{\int_{s}^{t}2C_{*}^{2}|\widetilde{v}(r)+\widetilde{z}^{\alpha}(r)|_{\tilde{H}}^{2}dr}\\
&  +\int_{s}^{t}\left(2 C_{*}^{2}|\widetilde{v}(r)+\widetilde{z}^{\alpha}(r)|_{\tilde{H}}^{2}%
\Vert\widetilde{z}^{\alpha}(r)\Vert_{\tilde{V}}^{2}+\frac{2\alpha^{2}}{\nu}|\widetilde{z}^{\alpha}%
(r)|_{\tilde{H}}^{2}\right)  e^{\int_{r}^{t}2C_{*}^{2}|\widetilde{v}(r^{\prime}%
)+\widetilde{z}^{\alpha}(r^{\prime})|_{\tilde{H}}^{2}dr^{\prime}}dr\,.
\end{align*}
Therefore, for $t=0$, we have%

\begin{align*}
\Vert\widetilde{v}(0)\Vert_{\tilde{V}}^{2} &  \leq\Vert\widetilde{v}(s)\Vert_{\tilde{V}}^{2}%
e^{\int_{s}^{0}2C_{*}^{2}|\widetilde{v}(r)+\widetilde{z}^{\alpha}(r)|_{\tilde{H}}^{2}dr}\\
&  +\int_{s}^{0}\left(2C_{*}^{2}|\widetilde{v}(r)+\widetilde{z}^{\alpha}(r)|_{\tilde{H}}^{2}%
\Vert\widetilde{z}^{\alpha}(r)\Vert_{\tilde{V}}^{2}+\frac{2\alpha^{2}}{\nu}|\widetilde{z}%
(r)|_{\tilde{H}}^{2}\right)  e^{\int_{r}^{0}2C_{*}^{2}|\widetilde{v}(r^{\prime}%
)+\widetilde{z}^{\alpha}(r^{\prime})|_{\tilde{H}}^{2}dr^{\prime}}dr\,.
\end{align*}
Now, we integrate over $(-1,0)$ to obtain%

\begin{eqnarray*}
\Vert\widetilde{v}(0)\Vert_{\tilde{V}}^{2}&\leq&\left(
\int_{-1}^{0}\Vert\widetilde
{v}(s)\Vert_{\tilde{V}}^{2}ds+\int_{-1}^{0}\left(
2C_{*}^{2}|\widetilde{v}(s)+\widetilde
{z}(s)^{\alpha}|_{\tilde{H}}^{2}\Vert\widetilde{z}^{\alpha}(s)\Vert_{\tilde{V}}^{2}+\frac{2\alpha^{2}}{\nu}%
|\widetilde{z}^{\alpha}(s)|_{\tilde{H}}^{2}\right)  ds\right)\\
&& \times \exp{\left(\int_{-1}^{0}2C_{*}^{2}%
|\widetilde{v}(s)+\widetilde{z}^{\alpha}(s)|_{\tilde{H}}^{2}ds\right)}\,.%
\end{eqnarray*}
Consequently, we use the estimate of the preceding lemma to complete
the proof.
\end{proof}

\begin{lemma} Let $\varphi^{\lambda}(t,\omega)$ be a stochastic
flow associated to equation (\ref{GOY_lambda}), defined on a
$\theta_{t}$-invariant full measure set $\Omega_{1}^{0}$. On the
$\theta_{t}$-invariant full measure set $\Omega_{W}^{0}$ described
previously, one can define $\widetilde{z}^{\alpha}(t,\omega)$ and
$\widetilde{v}(t,\omega,\widetilde{u}_{0})$, say for a given fixed
$\alpha =\alpha_{0}:= 2 \alpha_*(\nu)$ (see (\ref{E-z})). On the
$\theta_{t}$-invariant full measure set
$\Omega^{0}=\Omega_{1}^{0}\bigcap \Omega_{W}^{0}$ we have

$$\varphi^{\lambda}(t,\omega)\widetilde{u}_{0}=
\widetilde{v}(t,\omega,\widetilde{u}_{0})+\widetilde{z}^{\alpha}(t,\omega),$$
and for all $\omega\in\Omega^{0}$, there exists a compact
absorbing set at time 0 in $\widetilde{H}$ for
$\varphi^{\lambda}(t,\omega)$.
\end{lemma}
\begin{proof}
We have proved in Lemma \ref{radius_estimate4}, that the ball $B(0,R_{3}(\omega)$ is an absorbing set at time 0
in $\widetilde{V}$, which is compact in $\widetilde{H}$. Hence, defining $K(\omega):=\left\{u\in \tilde{V}\ :\
\|u\|_{\tilde{V}}^{2}\leq R_{3}+\|\widetilde{z}^{\alpha}(0,\omega)\|_{\tilde{V}}^{2}\right\}$ concludes the proofs.
\end{proof}

\begin{theorem}
For every value of the parameter $\lambda\in \mathbb{R}$, the random
dynamical system $\varphi^{\lambda}$ associated to the equation
(\ref{GOY_lambda}) has a unique global random attractor
$\mathcal{A}_{\lambda}(\omega)$.
\end{theorem}
\begin{proof}
Using Theorem \ref{thm_flandoli_crauel} and the existence of a compact absorbing set in
$\widetilde{H}$, we have the existence of a random attractor $\mathcal{A}%
_{\lambda}(\omega)$ which is forward invariant.
\end{proof}

We can now apply Theorem 2 from \cite{CarabLangaRob}. The statement of this
theorem is composed of two parts, the first one devoted to the convergence of
the random attractor to the deterministic one as the intensity of the noise
goes to zero; the second one to the upper semicontinuity of the random
attractor when the parameter of the noise varies with continuity to some
non-zero value. We apply the second part. The assumptions of the second part
are: i) the existence of the random attractor for every fixed value of the
parameter, ii) the $P$-a.s. continuous dependence of trajectories on the
parameter, in any fixed finite interval of time, uniformly in the initial conditions
taken from any fixed non-random bounded set.
Both assumptions have been proved in the previous sections. Thus
we get the following final result.

\begin{theorem} \label{theorem_upper_semicontinous}
Let $\mathcal{A}_{\lambda}\left(\omega\right)$ be the random attractor associated with equation
\eqref{GOY_lambda}, then
there is upper semicontinuous convergence of $\mathcal{A}_{\lambda}\left(
\omega\right)  $ to $\mathcal{A}_{0}\left(  \omega\right)  $ as $\lambda
\rightarrow0$:
\[
\lim_{\lambda\rightarrow0}d_{H}\left(  \mathcal{A}_{\lambda}\left(
\omega\right)  ,\mathcal{A}_{0}\left(  \omega\right)  \right) =0\,
\, \text{ with }\,\,  P- \text{a.s.}\, .%
\]
\end{theorem}

\subsection{Random squeezing property}

In this section, we are going to establish that the random attractor
of the random dynamical system $\varphi$ associated with equation
\eqref{GOY_lambda} has a finite
Hausdorff dimension (notice here that for simplicity of notation, we dropped the superscript ${\lambda}$ in $\varphi^{\lambda}$.
Let $\widetilde{u}$ and $\widetilde{v}$ be
two solutions of the associated equation (\ref{GOY_lambda}), then
the difference $\widetilde{u}-\widetilde{v}$ is
solution of%

\begin{equation}\label{difference_equation}
\frac{d(\widetilde{u}-\widetilde{v})}{dt}+\nu\widetilde{A}(\widetilde
{u}-\widetilde{v})= -\widetilde{B}_{\lambda}(\widetilde{u},\widetilde
{u}-\widetilde{v}) -\widetilde{B}_{\lambda}(\widetilde{u}-\widetilde
{v},\widetilde{v}).
\end{equation}

\begin{lemma}\label{lemma_squeezing1}
Let $\Pi$ be the orthogonal projection on the first $n$ eigenvectors of the operator $\widetilde{A}$. Then,

\begin{equation}\label{Hausdorff_1}
|\Pi\left(  \varphi(t,\omega)\widetilde{u}_{0}-\varphi(t,\omega)\widetilde
{v}_{0}\right)  |_{\tilde{H}}\leq|\widetilde{u}_{0}-\widetilde{v}_{0}|_{\tilde{H}}%
e^{\frac{C_{*}}{\nu}\displaystyle\int_{{0}}^{t}R_{1}(\theta_{s}\omega)ds}%
\end{equation}%

\begin{eqnarray}\label{Hausdorff_2}
&&|(I-\Pi)\left( \varphi(t,\omega)\widetilde{u}_{0}-\varphi(t,\omega
)\widetilde{v}_{0}\right)
|_{\tilde{H}}\leq\\
&&\hskip -0.35in
|\widetilde{u}_{0}-\widetilde{v}_{0}|_{\tilde{H}}\left( e^{-k_{n+1}\nu
t}+\left(\frac{\sqrt{2}C^{2}}{\left(\nu k_{n+1}\right)^{3/2}}\right)e^{\displaystyle
\int_{0}^{t}\left[R_{1}(\theta_{s}(\omega))\right]^{2}+
\frac{C_{*}}{\nu}R_{1}(\theta_{s}(\omega))ds}\right),\nonumber
\end{eqnarray}
where $C_{*}$ is the constant in inequality \eqref{V-prime}, and $C$ is the constant in the inequalities in Lemma \ref{lemma su B tilde},
for all $t\geq0$ and all $\widetilde{u}_{0},\widetilde{v}_{0}\in
\mathcal{A}(\omega)$, where $R_{1}=R_{1}^{\{t=0\}}$ given in Lemma
\ref{radius_estimate2},  say for a given fixed   $\alpha=\alpha_{0}:=
2\alpha_*(\nu)$ (see (\ref{E-z})).
\end{lemma}

\begin{proof}
We multiply equation (\ref{difference_equation}) by
$\widetilde{u}-\widetilde{v}$, and use inequality \eqref{V-prime} to obtain

\begin{align*}
\frac{1}{2}\frac{d}{dt}|(\widetilde{u}-\widetilde{v})|_{\tilde{H}}^{2}+\nu\Vert
(\widetilde{u}-\widetilde{v})\Vert_{\tilde{V}}^{2} &
\leq|<(\widetilde{B}_{\lambda
}(\widetilde{u}-\widetilde{v},\widetilde{v})),(\widetilde{u}-\widetilde
{v})>|\\
&
\leq\frac{\nu}{2}\Vert(\widetilde{u}-\widetilde{v})\Vert_{\tilde{V}}^{2}+\frac{1}{2\nu}\Vert
\widetilde{B}_{\lambda}(\widetilde{u}-\widetilde{v},\widetilde{v}%
)\Vert_{\widetilde{V}^{\prime}}\\
&
\leq\frac{\nu}{2}\Vert(\widetilde{u}-\widetilde{v})\Vert_{\tilde{V}}^{2}+\frac{C_{*}}{2\nu}|\widetilde
{u}-\widetilde{v}|_{\tilde{H}}^{2}|\widetilde{v}|_{\tilde{H}}^{2}.%
\end{align*}
Using Gronwall lemma, we obtain%

\[
|\widetilde{u}(t)-\widetilde{v}(t)|_{\tilde{H}}^{2}\leq|\widetilde{u}(t_{0})-\widetilde
{v}(t_{0})|_{\tilde{H}}^{2}e^{\frac{C_{*}}{\nu}\int_{t_{0}}^{t}|\widetilde{v}(s)|_{\tilde{H}}^{2}ds}%
\]
Now, using the invariance of the attractor, if we take $\widetilde{u}_{0},\widetilde{v}_{0}\in\mathcal{A}(\omega)$
then $\widetilde{u}(t),\widetilde{v}(t)\in\mathcal{A}(\theta_{t}\omega)$, %
and therefore by Lemma \ref{radius_estimate1} and Lemma
\ref{radius_estimate2} we have
\begin{equation}\label{(u-v)_squeezing1}
|\varphi(t,\omega)\widetilde{u}_{0}-\varphi(t,\omega)\widetilde{v}_{0}%
|_{\tilde{H}}^{2}\leq|\widetilde{u}_{0}-\widetilde{v}_{0}|_{\tilde{H}}^{2}e^{\frac{C_{*}}{\nu}\int_{{0}}^{t}%
R_{1}(\theta_{s}\omega)ds}%
\end{equation}

for all $t\geq0$ and all $\widetilde{u}_{0},\widetilde{v}_{0}\in
\mathcal{A}(\omega)$.

Recall that $\Pi$ is a projection on the n-dimensional subspace of eigenvectors of the operator $\widetilde{A}$,
we have
\[
|\Pi\left(
\varphi(t,\omega)\widetilde{u}_{0}-\varphi(t,\omega)\widetilde
{v}_{0}\right)
|_{\tilde{H}}^{2}\leq|\varphi(t,\omega)\widetilde{u}_{0}-\varphi
(t,\omega)\widetilde{v}_{0}|_{\tilde{H}}^{2}.
\]
Let $Q:=I-\Pi$, it commutes with $\widetilde{A}$ but not with $\widetilde
{B}_{\lambda}$.\\
Let us apply the operator $Q$ to the equation (\ref{difference_equation}),
then using Lemma \ref{lemma su B tilde} and the Poincare inequality we get%

\begin{align*}
\frac{1}{2}\frac{d}{dt}|Q(\widetilde{u}-\widetilde{v})|_{\tilde{H}}^{2}+\nu\Vert
Q(\widetilde{u}-\widetilde{v})\Vert_{\tilde{V}}^{2} &
\leq|<Q(\widetilde{B}_{\lambda
}(\widetilde{u},\widetilde{u}-\widetilde{v})),Q(\widetilde{u}-\widetilde
{v})>|\\
+| &  <Q(\widetilde{B}_{\lambda}(\widetilde{u}-\widetilde{v},\widetilde
{v})),Q(\widetilde{u}-\widetilde{v})>|\\
& \hskip -2in \leq
\left(|\widetilde{B}_{\lambda}(\widetilde{u},\widetilde{u}-\widetilde{v}%
)|_{\widetilde{H}}+|\widetilde{B}_{\lambda}(\widetilde
{u}-\widetilde{v},\widetilde{v})|_{\widetilde{H}}\right) |Q(\widetilde{u}-\widetilde{v})|_{\widetilde{H}}\\
& \hskip -2in\leq \frac{C}{\sqrt{k_{n+1}}}|\widetilde{u}-\widetilde{v}|_{\widetilde{H}}\left(\|\widetilde{u})\|_{\widetilde{V}}
+\|\widetilde{v})\|_{\widetilde{V}}\right)\|Q(\widetilde{u}-\widetilde{v})\|_{\widetilde{V}}\\
& \hskip -2in \leq\frac{\nu}{2}\Vert Q(\widetilde{u}-\widetilde{v})\Vert_{\tilde{V}}^{2}%
+\frac{C^{2}}{2k_{n+1}\nu}|\widetilde{u}-\widetilde{v}|_{\tilde{H}}^{2}\left(
|\widetilde{u}|_{\tilde{V}}^{2}+|\widetilde
{v}|_{\tilde{V}}^{2}\right).
\end{align*}

Now, using the Poincare inequality on the left side of the above inequality we get that

\begin{align*}
\frac{d}{dt}|Q(\widetilde{u}-\widetilde{v})|_{\tilde{H}}^{2}+
\nu k_{n+1}|
Q(\widetilde{u}-\widetilde{v})|_{\tilde{H}}^{2} &\leq
\frac{C^{2}}{k_{n+1}\nu}|\widetilde{u}-\widetilde{v}|_{\tilde{H}}^{2}\left(
|\widetilde{u}|_{\tilde{V}}^{2}+|\widetilde
{v}|_{\tilde{V}}^{2}\right).
\end{align*}
Hence,
\begin{eqnarray*}
 |Q(\widetilde{u}-\widetilde{v})(t)|^{2}_{\tilde{H}}\leq&|Q(\widetilde{u}-\widetilde{v}%
)(t_{0})|^{2}_{\tilde{H}}-k_{n+1}\nu\displaystyle\int_{t_{0}}^{t}|Q(\widetilde{u}-\widetilde{v}%
)(s)|^{2}_{\tilde{H}}ds\\
&+\frac{C^{2}}{k_{n+1}\nu}\displaystyle\int_{t_{0}}^{t}|(\widetilde{u}-\widetilde{v})(s)|_{\tilde{H}}^{2}\left(
|\widetilde{u}(s)|_{\tilde{V}}^{2}+|\widetilde{v}(s)|_{\tilde{V}}^{2}\right)
ds
\end{eqnarray*}
Using Gronwall lemma we get%

\begin{align*}
|Q(\widetilde{u}-\widetilde{v})(t)|^{2}_{\tilde{H}}&\leq|Q(\widetilde{u}-\widetilde{v}%
)(t_{0})|^{2}_{\tilde{H}}e^{-k_{n+1}\nu(t-t_{0})}\\
&+\frac{C^{2}}{k_{n+1}\nu}\int_{t_{0}}^{t}e^{-k_{n+1}\nu(t-s)}%
|(\widetilde{u}-\widetilde{v})(s)|_{\tilde{H}}^{2}\left(  |\widetilde{u}(s)|_{\tilde{H}}^{2}%
+|\widetilde{v}(s)|_{\tilde{H}}^{2}\right)  ds.
\end{align*}
Let us take $t_{0}=0$ and $\widetilde{u}_{0},\widetilde{v}_{0}\in
\mathcal{A}(\omega)$ , then we have, thanks to %
Lemma \ref{radius_estimate1} and Lemma \ref{radius_estimate2}, that%

\[
|Q(\widetilde{u}-\widetilde{v})(t)|^{2}_{\tilde{H}}\leq|\widetilde{u}_{0}-\widetilde{v}%
_{0}|^{2}_{\tilde{H}}e^{-k_{n+1}\nu t}+\frac{C^{2}}{k_{n+1}\nu}\int_{0}^{t}e^{-k_{n+1}\nu(t-s)}R_{1}(\theta_{s}%
(\omega))|(\widetilde{u}-\widetilde{v})(s)|_{\tilde{H}}^{2}ds.
\]
We use (\ref{(u-v)_squeezing1}) in the above inequality to obtain
\begin{align*}
|Q(\widetilde{u}-\widetilde{v})(t)|^{2}_{\tilde{H}} &
\leq|\widetilde{u}_{0}-\widetilde
{v}_{0}|^{2}_{\tilde{H}}\left(  e^{-k_{n+1}\nu t}+\frac{C^{2}}{k_{n+1}\nu}\int_{0}^{t}e^{-k_{n+1}\nu(t-s)}R_{1}%
(\theta_{s}(\omega))e^{\int_{0}^{s}\frac{C_{*}}{\nu}R_{1}(\theta_{r}(\omega))dr}ds\right)  \\
&  \leq|\widetilde{u}_{0}-\widetilde{v}_{0}|^{2}_{\tilde{H}}\left(
e^{-k_{n+1}\nu t}+\frac{C^{2}}{k_{n+1}\nu}\left(
e^{\int_{0}^{t}\frac{C_{*}}{\nu}R_{1}(\theta_{s}(\omega))ds}\right)  \int_{0}^{t}e^{-k_{n+1}%
\nu(t-s)}R_{1}(\theta_{s}(\omega))ds\right) \,.
\end{align*}
On the other hand  using Cauchy-Schwarz inequality and the fact that
$\sqrt{x}\leq
e^{x}$ for all $ x>0$,%

\begin{align*}
\displaystyle\int_{0}^{t}e^{-k_{n+1}\nu(t-s)}R_{1}(\theta_{s}(\omega))ds &
\leq\left(  \int_{0}^{t}e^{-2k_{n+1}\nu(t-s)}\right)  ^{1/2}\left(  \int_{0}%
^{t}\left[R_{1}(\theta_{s}(\omega))\right]^{2}ds\right)  ^{1/2}\\
&  \leq\left(  \frac{1}{2k_{n+1}\nu}\right)  ^{1/2}e^{\displaystyle\int_{0}%
^{t}\left[R_{1}(\theta_{s}(\omega))\right]^{2}ds}\,.%
\end{align*}

Combining all the above estimates we get that
\[
|Q(\widetilde{u}-\widetilde{v})(t)|^{2}_{\tilde{H}}\leq|\widetilde{u}_{0}-\widetilde{v}%
_{0}|^{2}_{\tilde{H}}\left(  e^{-k_{n+1}\nu t}+\frac{\sqrt{2}C^{2}}{\left(k_{n+1}\nu\right)^{3/2}}
e^{\displaystyle\int_{0}^{t}\left[R_{1}(\theta_{s}(\omega))\right]^{2}+\frac{C_{*}}{\nu}R_{1}%
(\theta_{s}(\omega))ds}\right)\,.
\]
\end{proof}

\subsection{Finite dimensionality of the random attractor}
In order to be able to apply Theorem \ref{theorem_debussche} we need
to show that $E(C_{\mathcal{H}} (\omega))<\infty,$ where
\begin{equation}\label{C_H}
C_{\mathcal{H}}(\omega):=\left[R_{1}(\omega)\right]^{2}+\frac{C_{*}}{\nu}R_{1}(\omega),
\end{equation}
is the exponent in equation (\ref{Hausdorff_2}) of the squeezing
Lemma \ref{lemma_squeezing1}. This is because $C_{\mathcal{H}}
(\omega)$ plays, in our case,  the role of  $C_5(\omega)$ in Theorem
\ref{theorem_debussche}. Here $R_1(\omega)=R_1^{\{t=0\}}(\omega)$
given in Lemma \ref{radius_estimate2} (see also Lemma
\ref{lemma_squeezing1}),  say for a given fixed $\alpha=
\alpha_{0}:=2\alpha_*(\nu)$ (see (\ref{E-z})).
In order to get the finite expectation of $C_{\mathcal{H}}$, i.e. $E(C_{\mathcal{H}}%
(\omega))<\infty,$ we need to estimate the moments of the radii
$R_{1}$.

\begin{proposition}\label{theorem_Z}
Let  $C_{*}>0$ be the constant in the inequality (\ref{V-prime}),
$\gamma_{0}=\nu k_0$, and $\alpha_0 = 2 \alpha_*(\nu)$ (see
(\ref{E-z})). Then the stationary process $\widetilde{z}_{t}$ that
solves  of the equation
\[
d\widetilde{z}_{t}=-\left(  \widetilde{A}+\alpha_0 \right)  \widetilde{z}%
_{t}^{{}}dt+d\widetilde{W}_{t}%
\]
satisfies
\[
E\left[  e^{C_{*}\int_{s}^{\tau}\left\| \widetilde{z}_{t}\right\|
_{\widetilde{V}}dt}\right]  \leq
\widetilde{C}_{2}e^{\gamma_{0}(\tau-s)},
\]
for all $s<\tau\le 0$, where $ \widetilde{C}_{2}=E\left[
e^{\frac{C_{*}^{2}\left\| \widetilde{z}_{0} \right\|
_{\widetilde{V}}^{2}}{4\gamma_{0}}}\right] < \infty$.
\end{proposition}

\begin{proof}
By Young's inequality we have
\[
C_{*}\left\|  \widetilde{z}_{t}\right\|  _{\widetilde{V}}=2\sqrt{\gamma_{0}%
(\tau-s)}\frac{C_{*}\left\|  \widetilde{z}_{t}\right\|  _{\widetilde{V}}}%
{2\sqrt{\gamma_{0}(\tau-s)}}\leq\gamma_{0}(\tau-s)+\frac{C_{*}^{2}\left\|
\widetilde {z}_{t}\right\|  _{\widetilde{V}}^{2}}{4\gamma_{0}(\tau-s)}\,,%
\]
therefore,
\[
E\left[  e^{C_{*}\int_{s}^{\tau}\left\|  \widetilde{z}_{t}\right\|
_{\widetilde{V}}dt}\right]  \leq e^{\gamma_{0}(\tau-s)}E\left[  e^{\frac{1}{(\tau-s)}%
\int_{s}^{\tau}\frac{C_{*}^{2}\left\|  \widetilde{z}_{t}\right\|
_{\widetilde {V}}^{2}}{4\gamma_{0}}dt}\right]  .
\]
Thus, it is sufficient to show that
\[
E\left[  e^{\frac{1}{(\tau-s)}\int_{s}^{\tau}\frac{C_{*}^{2}\left\|  \widetilde{z}%
_{t}\right\|  _{\widetilde{V}}^{2}}{4\gamma_{0}}dt}\right]  \leq
\widetilde{C}_{2}.
\]
Thanks to  Jensen inequality we have
\[
E\left[  e^{\frac{1}{(\tau-s)}\int_{s}^{\tau}\frac{C_{*}^{2}\left\|  \widetilde{z}%
_{t}\right\|  _{\widetilde{V}}^{2}}{4\gamma_{0}}dt}\right]  \leq\frac{1}%
{(\tau-s)}\int_{s}^{\tau}E\left[  e^{\frac{C_{*}^{2}\left\|
\widetilde{z}_{t}\right\| _{\widetilde{V}}^{2}}{4\gamma_{0}}}\right]
dt.
\]
Since $\widetilde{z}_{t}$ is a stationary process then it follows
that
\[
E\left[  e^{\frac{C_{*}^{2}\left\|  \widetilde{z}_{t}\right\|
_{\widetilde {V}}^{2}}{4\gamma_{0}}}\right]  =E\left[
e^{\frac{C_{*}^{2}\left\| \widetilde{z}_{0}\right\|
_{\widetilde{V}}^{2}}{4\gamma_{0}}}\right] < \infty \,,
\]
thanks to (\ref{E-z}). The proof is complete.
\end{proof}

Let us recall that

$$\displaystyle
R_{1}^{t}(\omega):=1+\int_{-\infty}^{t}f(s)e^{\int_{s}^{t}\left(
2C_{*}\Vert\widetilde{z}(r)\Vert-\frac{k_{0}\nu}{2}\right)dr}ds$$
where
\[
f(t):=\frac{4C^{2}_{*}}{\nu}|\widetilde{z}(t)|^{4}+\frac{8\alpha^{2}}{k_{0}%
\nu}|\widetilde{z}(t)|^{2}\,,
\]
and that $R_{1}(\omega)=R_{1}^{\{t=0\}}(\omega)$, which, as it has
been remarked in section \ref{absorbing-set}, depend on the
parameter $\alpha$. Hereafter we choose
$\alpha=\alpha_0=2\alpha_*(\nu)$ (see (\ref{E-z}) and Propositon
\ref{theorem_Z}). Therefore, the relevant results of section
\ref{absorbing-set} are valid for this choice of $\alpha$.



\begin{lemma} Let $\gamma_{0}$ and $\alpha_0$ be as in Proposition
\ref{theorem_Z}. Then
\[
E((R_{1}(\omega))^{2})<\infty,
\]
and by the Cauchy-Schwarz inequality
\[
E R_{1}(\omega)<\infty.
\]
Consequently,
\begin{equation}\label{average_C_H}
E\left(C_{\mathcal{H}}\right)<\infty,
\end{equation}
where $C_{\mathcal{H}}(\omega)$ is given in equation (\ref{C_H}).
\end{lemma}

\begin{proof}
First we observe that using successively Jensen inequality, Fubini
Theorem and H\"{o}lder inequality yield
\begin{align*}
E(R_{1}(\omega))^{2} &  =E\left(
1+\int_{-\infty}^{0}f(s)e^{\displaystyle
\int_{s}^{0}\left(  2C_{*}\Vert\widetilde{z}(r)\Vert-\frac{k_{0}\nu}%
{2}\right)  dr}ds\right)  ^{2}\\
&  \leq2+E\left(
\int_{-\infty}^{0}f(s)e^{\displaystyle\int_{s}^{0}\left(
2C_{*}\Vert\widetilde{z}(r)\Vert-\frac{k_{0}\nu}{2}\right)
dr}ds\right)
^{2}\\
&
\leq2+E\int_{-\infty}^{0}f(s)^{2}e^{\displaystyle2\int_{s}^{0}\left(
2C_{*}\Vert z(r)\Vert-\frac{k_{0}\nu}{2}\right)  dr}ds\\
&
\leq2+\int_{-\infty}^{0}E\left(f(s)^{2}e^{\displaystyle2\int_{s}^{0}\left(
2C_{*}\Vert\widetilde{z}(r)\Vert-\frac{k_{0}\nu}{2}\right)  dr}\right)ds\\
&  \leq2+\int_{-\infty}^{0}\left(  Ef(s)^{4}\right)  ^{1/2}\left(
Ee^{\displaystyle4\int_{s}^{0}\left(
2C_{*}\Vert\widetilde{z}(r)\Vert -\frac{k_{0}\nu}{2}\right)
dr}\right)  ^{1/2}ds.
\end{align*}%

The process $\widetilde{z}$ is stationary, thus
\begin{equation*}
E(f(s)^{4})=E(f(0)^{4}).
\end{equation*}
All moment of a Gaussian random variable are finite, hence

\begin{equation*}
E(f(0)^{4})<\infty
\end{equation*}

Now, using the preceding estimates and Proposition \ref{theorem_Z}
we conclude that

\begin{align*}
E(R_{1}(\omega))^{2} &  \leq2+\left(  Ef(0)^{4}\right)
^{1/2}\int_{-\infty}^{0}\left( Ee^{\displaystyle
\int_{s}^0\left(  8C_{*}\Vert\widetilde{z}(r)\Vert-2k_{0}\nu%
\right)  dr}\right)  ^{1/2}ds\\
&  \leq2+\left(  Ef(0)^{4}\right) ^{1/2}
\int_{-\infty}^{0}e^{\displaystyle k_{0}\nu s}\left(
Ee^{\displaystyle\int_{s}^{0}8C_{*}\Vert\widetilde{z}(r)\Vert
dr}\right) ^{1/2}ds\\
& \leq2+\sqrt{\widetilde{C}_{2}}\left(  Ef(0)^{4}\right) ^{1/2}
\int_{-\infty}^{0}e^{\displaystyle
(k_{0}\nu-\frac{\gamma_{0}}{2})s}ds.
\end{align*}
Since $ \gamma_{0}= \nu k_{0}$, then
\[
E((R_{1}(\omega))^{2}) <\infty,
\]
and the proof is complete.
\end{proof}

As a consequence, we have the following theorem

\begin{theorem} Let $K_{1}$, $K_{2}$ and $K_{3}$ be the
absolute constant stated in Theorem \ref{theorem_debussche}. Let $n$
be large enough such that

\begin{equation*}
\frac{\sqrt{2}C^{2}}{\left(\nu k_{n+1}\right)^{3/2}}\leq K_{1},\ \ \
{ and}\ \  k_{n+1}\nu\geq K_{2}E\left(C_{\mathcal{H}}\right),
\end{equation*}
where $C_{\mathcal{H}}$ is given in \eqref{C_H}, for
$\alpha=\alpha_0=\alpha_*(\nu)$ (see (\ref{E-z})). Then, $P$-a.s.
the random attractor $\mathcal{A}_{\lambda}(\omega)$ of the random
dynamical system $\varphi^{\lambda}$ associated with equation
\eqref{GOY_lambda} has finite Hausdorff dimension which is less than
$K_{3}n\ln n$.
\end{theorem}
\begin{proof}
The proof follows from applying Theorem \ref{theorem_debussche} for
$\mu=\frac{\sqrt{2}C^{2}}{\left(\nu k_{n+1}\right)^{3/2}}$ and
$\delta=k_{n+1}\nu$. Then by virtue of \eqref{average_C_H} all the
assumptions of Theorem \ref{theorem_debussche} are satisfied. Hence,
we get that $P$-a.s. the random attractor
$\mathcal{A}_{\lambda}(\omega)$ of the random dynamical system
$\varphi^{\lambda}$ associated with equation \eqref{GOY_lambda} has
finite Hausdorff dimension which is less than $K_{3}n\ln n$. This
completes the proof.
\end{proof}

\vskip 025in

 \textbf{Acknowledgment:} The work of H. Bessaih was
supported in part  by the NSF grand No. DMS-0608494. The work of
E.S. Titi was supported in part by the NSF grant No. DMS-0708832,
and the ISF grant No. 120/6.

\end{document}